\documentclass[prb, twocolumn, superscriptaddress, aps, longbibliography,
floatfix, reprint]{revtex4-2}

\usepackage[utf8]{inputenc}
\usepackage[T1]{fontenc}

\usepackage{graphicx} 
\usepackage{amsmath}  
\usepackage{amssymb}
\usepackage{amsfonts}
\usepackage{dcolumn}
\usepackage{bm}    
\usepackage[]{xcolor}
\usepackage[]{newtxtext} 
\usepackage{pgffor}
\usepackage[subscriptcorrection,nosymbolsc,smallerops,bigdelims]{newtxmath} 
\usepackage[]{placeins} 
\usepackage{hyperref}  
\hypersetup{
	colorlinks,
	linkcolor={blue!90!black}, 
	citecolor={blue!90!black},
	urlcolor=	{blue!90!black}
}

\DeclareMathAlphabet{\mathcal}{OMS}{cmsy}{m}{n}
\DeclareMathAlphabet{\mathbcal}{OMS}{cmsy}{b}{n}

\newcommand{\kp}{\mbox{$\bm{k}{\cdot}\bm{p}$}}

\usepackage{pdfpages}

\makeatletter
\AtBeginDocument{%
  \let\LS@rot\@undefined   % ← to jest kluczowa linia
}
\makeatother

\begin{document}

\title{Coupling Quantum Dots to Elastic Waves in a Phononic Crystal Waveguide}

\author{Jakub Rosi\'n{}ski}
\email{jakub.rosinski@pwr.edu.pl}
\affiliation{Institute of Theoretical Physics, Wroc\l{}aw University of Science and Technology, Wroc\l{}aw, 50-370, Poland}

\author{Micha\l{} Gawe\l{}czyk}
\affiliation{Institute of Theoretical Physics, Wroc\l{}aw University of Science and Technology, Wroc\l{}aw, 50-370, Poland}

\author{Matthias  Wei\ss}
\affiliation{Institute of Physics, University of M\"unster, 48149, M\"unster, Germany}
\author{Hubert J. Krenner}
\affiliation{Institute of Physics, University of M\"unster, 48149, M\"unster, Germany}
\author{Pawe\l{} Machnikowski}
\email{pawel.machnikowski@pwr.edu.pl}
\affiliation{Institute of Theoretical Physics, Wroc\l{}aw University of Science and Technology, Wroc\l{}aw, 50-370, Poland}

\begin{abstract}
We present a comprehensive study of quantum dot (QD) coupling to various phononic modes in a phononic waveguide, combining multiband \kp{} and configuration-interaction (CI) QD state simulations with finite-element waveguide mode modeling. We consider self-assembled Stranski-Krastanov InGaAs/GaAs as well as local droplet-etched GaAs/AlGaAs structures. 
 Using \kp{}-CI calculations, we quantify the strain and piezoelectric responses of InAs and GaAs QDs. By systematically isolating volumetric/shear deformation-potential and piezoelectric channels, we demonstrate how mode symmetries dictate distinct coupling mechanisms. We identify the dominant coupling channels and characterize their observable signatures in the QD response. We predict strong linear energy shifts under volumetric strain and quadratic behavior under shear strain, especially in GaAs QDs. The piezoelectric effect is dominated by polarizability, which also leads to a quadratic response. The simulations show energy modulations up to 0.7~meV for an acoustic wave with 0.1~nm amplitude. The quadratic response to shear strain and piezoelectric field leads to frequency doubling in the QD response to a mechanical wave and to non-harmonic time traces when linear and quadratic effects contribute to a similar degree. The deep understanding of QD-acoustic couplings opens pathways to the optimal design of QD and waveguide structures, as well as to improved engineering of acousto-optic quantum interfaces.
\end{abstract}

% \date{\today}

\maketitle

\section{Introduction} 
Elastic waves and acoustic phonons in the form of radio-frequency Rayleigh surface acoustic waves (SAWs)~\cite{rayleigh1885waves} or Lamb waves~\cite{lamb1917waves} can interact with virtually any excitation within condensed matter~\cite{delsing20192019}. This universal coupling makes them exceptionally well suited for designing and realizing hybrid quantum systems~\cite{kurizki2015quantum, smith2016editorial} and quantum transduction~\cite{schutz2017universal, lauk2020perspectives}. An additional advantage of acoustic wave devices is their operation at gigahertz frequencies, allowing them to reach the phononic ground state without requiring active cooling protocols~\cite{satzinger2018quantum}. Radio-frequency elastic waves have been utilized to effectively control a wide range of quantum systems. Prominent examples include superconducting qubits at the single-phonon level~\cite{gustafsson2014propagating}, optically active quantum dots (QDs)~\cite{gell2008modulation, volk2010enhanced, metcalfe2010resolved, blattmann2014entanglement, schulein2015fourier, weiss2018interfacing,weiss2021optomechanical, wigger2021remote,choquer2022quantum}, single spins~\cite{whiteley2019spin, maity2020coherent}, defect centers~\cite{golter2016optomechanical, lazic2019dynamically, whiteley2019spin, hernandez2021acoustically}, and two-dimensional quantum emitters such as monolayer WSe$_2$ and h-BN~\cite{patel2024surface}. Moreover, SAWs facilitate on-chip transfer of quantum states between superconducting qubits using individual SAW quanta as the carriers of quantum information~\cite{bienfait2019phonon} or transfer single-electron spins between electrostatic QDs~\cite{hermelin2011electrons, mcneil2011demand,takada2019sound,jadot2021distant}. 

Given the broad applicability of acoustic phonons, there is a growing importance in mastering the manipulation of ultrahigh-frequency acoustic waves to advance phononic technology. Here, structures based on phononic crystals (PnC)~\cite{narayanamurti1979selective, martinez1995sound, benchabane2006evidence} offer promising routes. PnCs are periodic composite structures in which the elastic properties exhibit a periodic variation in space. By using this engineered structure, the dispersion relation and band gap of acoustic waves can be tailored~\cite{sigalas1993band}, thereby enabling control over acoustic propagation. For example, by perturbing the periodic geometry, one can generate a cavity~\cite{mohammadi2009high} or a waveguide~\cite{benchabane2011observation, otsuka2013broadband, ghasemi2018acoustic, pourabolghasem2018waveguiding} supported by the band gap, effectively trapping or guiding acoustic waves within specific spatial regions. Additionally, it is possible to control the propagation of elastic waves by slowing them down through a waveguide, further enhancing the manipulation capabilities of the system~\cite{modica2020slow}. Regarding the scalability of device integration, utilizing a suspended PnC membrane proves to be an appropriate platform. This is due to its capability to achieve both a high quality factor and a small mode volume, resulting in minimal loss and a compact device footprint. Additionally, it allows the incorporation of various components into a two-dimensional membrane, enabling intricate phonon manipulations. 

Optically active epitaxial QDs offer unique advantages in designing hybrid quantum architectures. First, semiconductor QDs offer significant advantage due to their inherent transitions naturally occurring within the visible and near-infrared spectral range, making them suitable for comprehensive realization of transduction into the optical frequency range. Second, their emission wavelength can be adjusted through chemical composition and size~\cite{garcia1998electronic}, or post-growth through external parameters such as electric or magnetic fields~\cite{bayer2002fine} or strain~\cite{trotta2012nanomembrane}. Moreover, QDs offer the versatility of integration within electrically active devices~\cite{drexler1994spectroscopy, krenner2006optically, salter2010entangled, zhang2015high, krenner2008semiconductor} and can be coupled to the dynamic strain associated with an elastic wave via the deformation potential (DP) mechanism or by the acoustic wave-generated electric fields in piezoelectric materials~\cite{weiss2018interfacing}.

The coupling between QDs and elastic waves has attracted considerable interest and is actively investigated experimentally. So far, experiments on strain-mediated coupling have been performed for Rayleigh SAWs~\cite{metcalfe2010resolved,weiss2018interfacing, weiss2021optomechanical,wigger2021resonance,choquer2022quantum,imany2022quantum} and Lamb waves~\cite{weiss2016surface, carter2017sensing,vogele2020quantum, spinnler2024quantum,spinnler2024single}. In addition, strain-mediated coupling has also been demonstrated using nanomechanical resonators operating in the MHz range~\cite{yeo2014strain,carter2018spin,yuan2019frequency,tanos2024high,carter2017sensing}.

In this paper, we investigate the coupling of QDs to dynamic elastic and piezoelectric fields co-propagating in a PnC waveguide. Specifically, we consider a PnC waveguide realized in a two-dimensional snowflake-type PnC~\cite{safavi2010design}, patterned into suspended GaAs and Al(Ga)As membranes to enable efficient elastic waveguiding and thereby facilitate strong interaction between phonons and the QD emitter. We characterize the coupling between acoustic-wave-induced strain fields and QD charge states via the conduction- and valence-band DPs, as well as via the strain-induced piezoelectric fields, both using a detailed numerical approach combining \kp{} and configuration-interaction (CI) methods and a simplified effective approach. We design a waveguide that enables operating frequencies within the gigahertz range, which is required for the resolved sideband regime~\cite{weiss2021optomechanical,choquer2022quantum}. Furthermore, we determine the optimum position of the QDs in the plane of the two-dimensional snowflake pattern and in the perpendicular direction relative to the center of the suspended membrane. Our simulations form the basis for the development of strategies to enhance the coupling between the optical emitter and tailored GHz-range elastic waves, which is a critical prerequisite for efficient quantum transduction between microwave and optical frequencies at technologically accessible temperatures~\cite{choquer2022quantum}.

This paper is organized as follows. In Sec.~\ref{sec:model}, we provide the theoretical background, starting with the PnC structures employed in our study (Sec.~\ref{subsec:crystal}). We then introduce the QD model used in our calculations (Sec.~\ref{subsec:kp_calc}), followed by the description of their coupling to elastic waves via strain and electric-field mechanisms (Sec.~\ref{sec:model-dp} and~\ref{sec:model-piezo}). In Sec.~\ref{sec:results}, we present the results, beginning with the band structure of the PnC structures (Sec.~\ref{sec:band}), followed by the QD response to external strain and electric fields (Sec.~\ref{sec:QDsuscept}). Finally, we discuss waveguide mode-dependent responses, including strain and electric-field profiles associated with different symmetries and the resulting QD response (Sec.~\ref{sec:QDResponse}). We conclude the paper in Sec.~\ref{sec:conclusions}. The Appendix provides additional information and data.

%%%%%%%%%%%%%%
\section{System, model and methods}\label{sec:model}

In this section, we describe the waveguide design and coupling mechanisms, taking into account the constraints imposed by realistic experimental conditions. 
The material constants used in the simulations are listed in Appendix~\ref{sec:mat-params}. 

\subsection{Phononic structures}\label{subsec:crystal}

\begin{figure}[tb]
\includegraphics[width=\columnwidth]{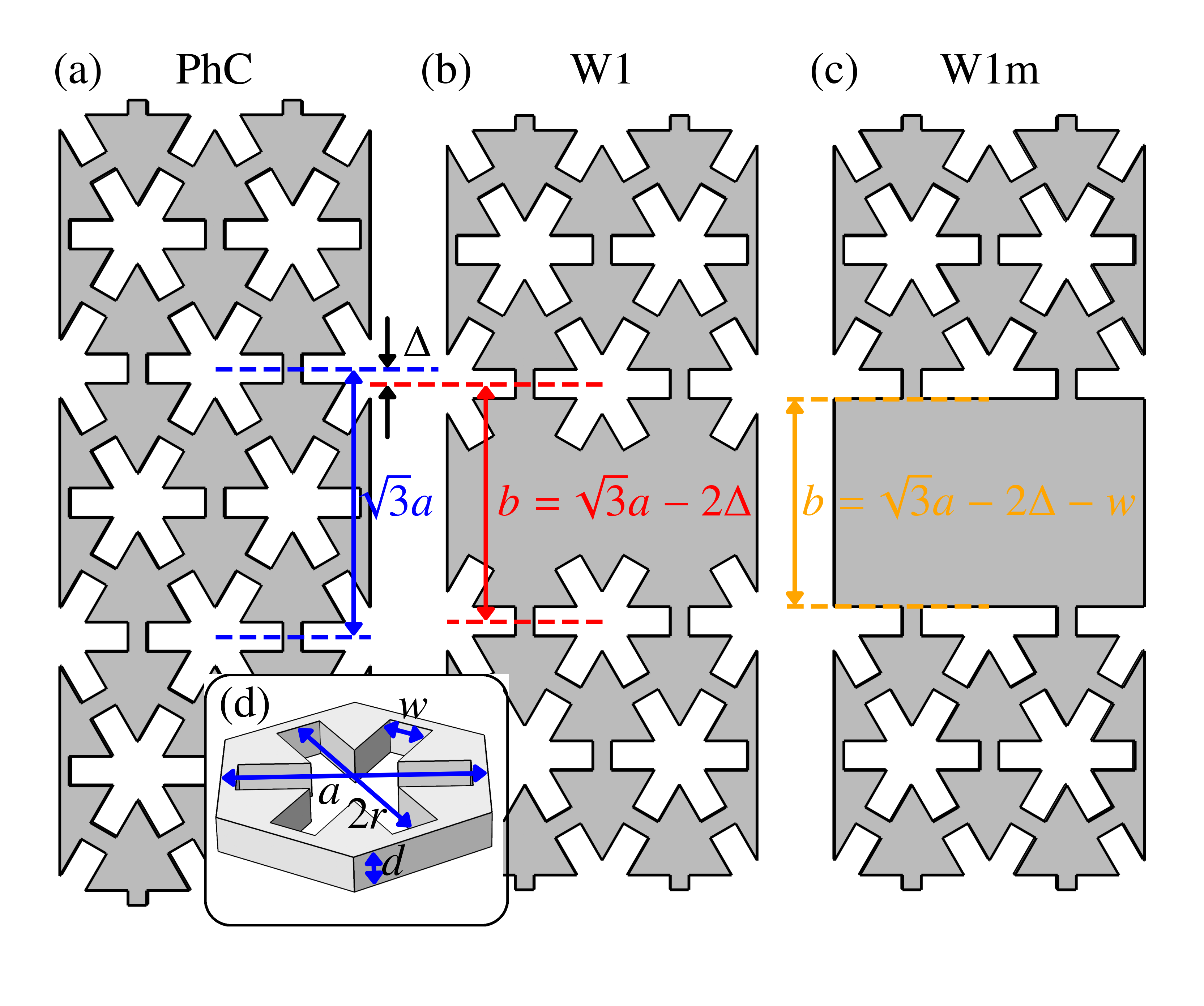}
    \caption{(a) PnC slab with a hexagonal snowflake lattice. (b) Standard W1 waveguide formed by removing a row of holes along the \( x \)-direction (in $\Gamma-K$ direction in reciprocal space) and shifting the adjacent lattice by \( \Delta \). (c) Modified waveguide W1m studied in this work, with two arms of the snowflakes additionally removed. 
    (d) Unit cell of the hexagonal snowflake PnC. }
    \label{fig:fig1-crystal-waveguide}
\end{figure}

The PnC structures considered here consist of hexagonal arrays of snowflake-shaped inclusions~\cite{safavi2010design, brendel2018snowflake, hatanaka2020real} in thin (001)-oriented GaAs and Al$_{0.4}$Ga$_{0.6}$As plates, as shown in Fig.~\ref{fig:fig1-crystal-waveguide}(a). The selected material platforms meet the integration requirements for the QDs to be considered in this work, and the two plate materials correspond to typical matrices for self-assembled InAs~\cite{Ardelt2016} and local-droplet-etched GaAs QDs~\cite{Yuan2023,schimpf2025optical}. Snowflake-shaped inclusions are chosen for their large number of adjustable parameters, such as the radius and width of each inclusion, enabling precise band-gap tuning. Additionally, the unique snowflake crystal structure, with large masses connected by a narrow bridge, creates a broad mechanical gap, whose width increases as the connecting bridges become narrower with increasing snowflake radius~\cite{safavi2010design}. The parameters governing this structure are denoted as $a$, $r$, $w$, and $d$, representing the lattice constant, radius, width of the snowflake inclusions, and thickness of the PnC structure, respectively [Fig.~\ref{fig:fig1-crystal-waveguide}(a,d)].  

One of the most common types of waveguides, referred to as the W1 waveguide, is illustrated in Fig.~\ref{fig:fig1-crystal-waveguide}(b). This particular waveguide is created by removing a row of snowflake-shaped holes in the $x$-direction, corresponding to the $\Gamma$-K direction in reciprocal space. The W1 waveguide can be modified by shifting the snowflake holes located above and below the waveguide toward the waveguide center by $\Delta$, resulting in a reduction of the effective waveguide core width by $2\Delta$ [Fig.~\ref{fig:fig1-crystal-waveguide}(b)]. Further modification leads to a structure shown in Fig.~\ref{fig:fig1-crystal-waveguide}(c) and is referred to as W1m. In this design, which is considered in this work, the two arms of the snowflake holes extending toward the waveguide core are additionally removed.

Elastic waves in a piezoelectric medium couple to charge distributions directly via strain fields, as well as by the strain-induced piezoelectric fields. The piezoelectric constitutive equations, which couple the mechanical and electrostatic degrees of freedom in a linear medium, have the form~\cite{auld1973acoustic}
\begin{subequations}
\begin{align}
    \label{eq:3a}
    &\tau_{ij}  = C^E_{ijkl}  \partial_l u_{k}  + e_{kij}  \partial_k \phi,\\
     \label{eq:3b}
    &\mathcal{D}_i  = e_{ikl} \partial_l u_{k}  - \kappa^S_{il} \partial_l \phi,
\end{align}
\end{subequations}
where $\tau_{ij}$ is the stress tensor, $C_{ijkl}^E$ is the elastic stiffness tensor at constant electric field, $u_i$ is the component of the displacement field, $e_{ikl}$ is the piezoelectric tensor, $\phi$ is the potential associated with the piezoelectric field, $\mathbcal{D}$ is the electric displacement field, and $\kappa^S_{ij}$ is the dielectric tensor at constant strain and at low frequency. Here and in the following, we use the Einstein summation convention. The equations of motion for elastic waves in a linear piezoelectric medium are obtained by applying Newton's law to the elastic medium, combined with Gauss's law~\cite{yudistira2012non}
\begin{equation}
    \label{eq:1}
    \rho \ddot{u}_i = \partial_j \tau_{ij} ,\quad\quad
     \partial_i D_{i}  = 0.
\end{equation}
Substituting Eq.~\eqref{eq:3a} and Eq.~\eqref{eq:3b} to Eq.~\eqref{eq:1}, one obtains a system of equations for $\bm{u}$ and $\phi$. These two functions are written in the form
\begin{equation}
\bm{u}(t) = \frac{1}{2}\tilde{\bm{u}}e^{-i\omega t} + \mathrm{c.c.}, \quad
\phi(t) = \frac{1}{2}\tilde{\phi}e^{-i\omega t} + \mathrm{c.c.} \label{eq:u}
\end{equation} 
and the resulting equations for the eigenmodes $\tilde{u}$ and $\tilde{\phi}$ are solved in the frequency domain by complete three-dimensional finite-element simulations using COMSOL Multiphysics~\cite{linkComsol}. 
From the displacements, we calculate the components of the strain tensor $\varepsilon_{ij}$ and define their respective complex amplitudes $\tilde{\varepsilon}_{ij}$ and phases $\varphi_{ij}$,
\begin{align}
\varepsilon_{ij}(t) &= 
\frac{1}{2}\left( \partial_i u_j(t) + \partial_j u_i(t) \right) = 
\frac{1}{2}\tilde{\varepsilon}_{ij}e^{-i\omega t} + \mathrm{c.c.} 
\nonumber \\
& = \left| \tilde{\varepsilon}_{ij} \right|\cos(\omega t -\varphi_{ij}),
\label{eq:strain-evol}
\end{align}
Similarly, the electric field components are derived from the spatial derivatives of the electric potential 
\begin{equation}\label{eq:Ei}
\mathcal{E}_i(t) = -\partial_i \phi(t) = \frac{1}{2}\tilde{\mathcal{E}}_i e^{-i\omega t} + \text{c.c.} = |\tilde{\mathcal{E}}_i|\cos(\omega t - \theta_i),
\end{equation}
where $\tilde{\mathcal{E}}_i$ and $\theta_i$ represent the complex amplitude and phase. We will also discuss field gradients, which are calculated and decomposed in the same way, with the complex amplitudes 
\begin{equation}
\tilde{\mathcal{E}}_{ij} = \partial_j\tilde{\mathcal{E}}_i = -\partial_j\partial_i \tilde{\phi} = |\tilde{\mathcal{E}}_{ij}|e^{-i\theta_{ij}},
\end{equation}
where $\theta_{ij}$ denotes the corresponding phase.
 \begin{figure}[tb]
	\includegraphics[width=\columnwidth]{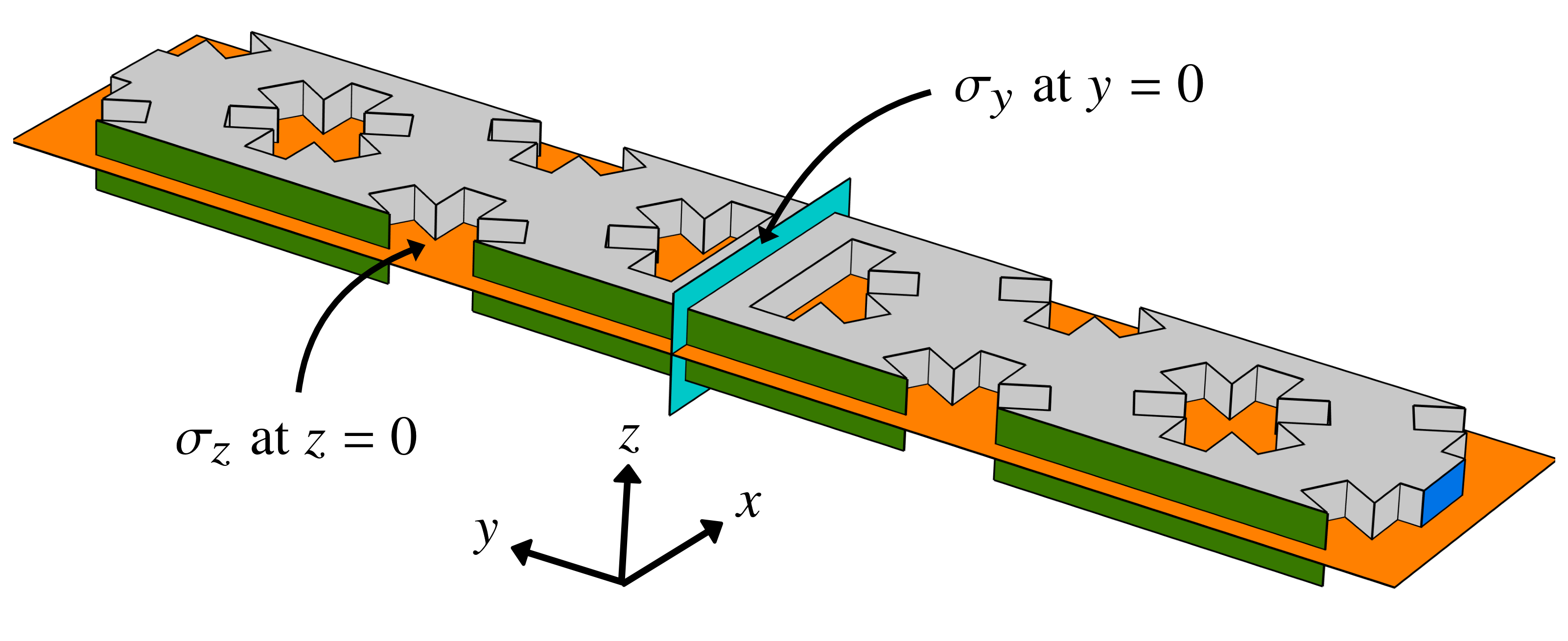}
    \caption{Supercell of the designed PnC waveguide with boundary conditions used in the simulation. Floquet and periodic boundary conditions are applied along the \(x\) and \(y\) directions, respectively (green and blue boundaries), while all other surfaces are traction-free (gray surfaces). Symmetry planes are shown at \(z=0\) (\(\sigma_z\) -- orange plane) and \(y=0\) (\(\sigma_y\) -- cyan plane).}
    \label{fig:fig2-supercell-boundary}
\end{figure}

For PnCs, a single unit cell is simulated [Fig.~\ref{fig:fig1-crystal-waveguide}(d)]. Floquet (Bloch) boundary conditions are applied to its three pairs of lateral faces in the $xy$ plane. Stress-free boundary conditions are applied to all remaining surfaces, which means that the normal stress components on the surfaces are zero. For the two boundary surfaces ($z=\pm d/2$) this gives
\[
\left. \tau_{3j}  \right|_{z=\pm d/2} = \left. C_{3jkl}  \partial_l u_k   \right|_{z=\pm d/2} = 0.
\]
For waveguides, we use a supercell as shown in Fig.~\ref{fig:fig2-supercell-boundary}, which is wide enough to ensure convergence for modes confined within the waveguide. 
Mechanical waves propagating in the waveguide are commonly generated electrically using interdigital transducers, which is efficient only along the [110] direction on a (001) surface~\cite{de2005modulation}, where the piezoelectric coupling is strong. Consequently, we assume the waveguide to be oriented in the $x$ direction as shown in Fig.~\ref{fig:fig2-supercell-boundary}, where $x$, $y$, and $z$ correspond to the $[110]$, $[\bar{1}10]$, and $[001]$ crystallographic directions, respectively.
Floquet boundary conditions are imposed in the propagation direction \(x\), periodic boundary conditions are used in the transverse direction \(y\), and stress-free conditions are applied to all remaining surfaces, as indicated by colors in Fig.~\ref{fig:fig2-supercell-boundary}.

A waveguide exhibits two planes of symmetry: the horizontal plane (\(\sigma_z\) at \(z=0\)) and the vertical plane (\(\sigma_y\) at \(y=0\)) [Fig.~\ref{fig:fig2-supercell-boundary}], allowing for the classification of its vibrational eigenmodes according to the corresponding symmetries. These classes are denoted as SS, SA, AS, and AA, where the first symbol indicates symmetry (S) or antisymmetry (A) with respect to the horizontal ($\sigma_z$) plane, and the second symbol refers to the symmetry or antisymmetry with respect to the vertical ($\sigma_y$) plane. In thin plates, these modes can be categorized into shear horizontal (SH) modes, characterized by linearly polarized displacement parallel to the plate surface and perpendicular to the propagation direction, and Lamb modes, which exhibit elliptical polarization with displacement components in the vertical plane. The Lamb modes can be further divided based on their symmetry with respect to the mid-plane of the plate (\(\sigma_z\)): symmetric Lamb modes (LS), often referred to as breathing modes, where the strain distribution is symmetric relative to the mid-plane, and antisymmetric Lamb modes (LA), known as flexural modes, where the strain is antisymmetric with respect to this plane. In PhC-based structures, these modes are coupled, which complicates their classification. In practice, however, the modes of the waveguide considered here largely retain predominant characteristics of one of the fundamental plate mode types. SA modes are predominantly SH-like, SS modes are predominantly LS-like, and AS/AA modes are predominantly LA-like. The weak mixing of the fundamental plate modes results from the smooth edges of the waveguide core in our geometry.

To make the polarization characteristics more quantitative, we define the average polarization component in the direction $l$, defined as
\begin{equation}\label{eq:Pl}
P_l = \dfrac{\int_{{\rm SC}} \left| \tilde{u}_l \right|^2 dV}{ \int_{{\rm SC}} \left| \tilde{\bm{u}} \right|^2  dV},    
\end{equation}
where \( \tilde{u}_{l} \) is a displacement component and SC denotes the integration over the volume of the supercell. Predominantly Lamb modes will show dominance of out-of-plane displacement (LA-like modes) or in-plane displacement along the propagation direction (LS-like modes). SH-like modes have mostly in-plane displacement, perpendicular to the propagation direction.

\subsection{Quantum dot modeling}\label{subsec:kp_calc}
To realistically calculate the coupling of QD excitonic states to acoustic modes, we model two types of application-relevant semiconductor QDs: self-assembled InAs/GaAs and GaAs/AlGaAs QDs grown by local droplet etching epitaxy. The InAs QDs are modeled as dome-shaped structures with 10\% in-plane asymmetry with a mean base diameter of 40~nm and 5~nm height, while the GaAs QD is modeled based on an atomic force microscopy scan of an actual nanohole with close to conical shape with 60~nm base diameter and 9~nm height~\cite{Yuan2023} (see Appendix~\ref{sec:mat-params} for details). QD models are discretized on a regular numerical mesh, after which we simulate material interdiffusion at interfaces by applying Gaussian averaging with a spatial extent of 0.9~nm. We calculate the structural (built-in) strain in such numerical QD models by minimizing the elastic energy within continuum elasticity theory. Subsequently, we calculate the shear strain-induced built-in (static) piezoelectric field with terms up to second order in strain tensor elements, which is needed for larger structural strain as opposed to a much weaker one caused by elastic waves, where keeping linear terms is sufficient.

Next, we find the electron, hole, and excitonic states by using a combination of multiband $\kp$~\cite{Bahder1992} and configuration-interaction~\cite{BryantPRL1987} methods. For this, we use a custom numerical code~\cite{Gawarecki2014,Gawarecki2018} that accounts for (structural and external) strain, piezoelectric field, and spin-orbit effects. We calculate 6 (12) electron and 6 (12) hole states for InAs (GaAs) QDs. The difference is due to the weaker confinement regime in the larger GaAs QDs, leading to lower level splittings and relatively stronger particle interaction effects. From these states, we form a basis of electron-hole configurations, in which the direct Coulomb and exchange interaction, including phenomenological electron-hole exchange, are then diagonalized to find the exciton eigenstates.

Such modeling allows us to directly calculate the exciton energy shift due to an arbitrary external strain tensor, which yields the strength of various DP couplings, as described in Sec.~\ref{sec:model-dp}. We can also decompose the actual exciton ground-state charge distribution into the multipole components needed to calculate the strength of the piezoelectric coupling, as discussed in Sec.~\ref{sec:model-piezo}. Finally, we can also calculate a total response via both couplings directly.

\subsection{Deformation potential couplings}
\label{sec:model-dp}

QDs interact with the dynamic strain associated with an elastic wave through the DP~\cite{krummheuer2005coupled, gell2008modulation, metcalfe2010resolved, schulein2015fourier, pustiowski2015independent, wigger2017systematic, nysten2017multi, weiss2018interfacing, wigger2021resonance}. We take into account both conduction-band and valence-band DPs, the latter including both volumetric, as well as shear-strain components.

The conduction-band DP coupling is proportional to the relative volume change $\delta V/V$ associated with a propagating Lamb wave, and can be expressed as~\cite{madelung1978introduction, krummheuer2005coupled}
\begin{equation}\label{eq:e2}
    \Delta E_{\mathrm{vol}}  = D_{\rm vol}  \dfrac{\delta V }{V},
\end{equation}
where $D_{\rm vol}$ is the exciton DP strength. As the relative volume change is equal to the volumetric strain, which is the trace of the strain tensor, we can rewrite Eq.~\eqref{eq:e2} as
\begin{equation}
    \Delta E_{\mathrm{vol}}  = D_{\rm vol} {\mathrm{Tr}} \varepsilon. \label{eq:Evol}
\end{equation}

Valence-band DPs have both volumetric and shear-strain contributions, as described by the Bir-Pikus Hamiltonian~\cite{yu2005fundamentals}, and can, therefore, lead to energy shifts in the absence of volumetric strain. 
Shear strain couples a heavy-hole subband to the two light-hole subbands with different spin orientations via terms $-d_{\mathrm{v}}(\varepsilon_{xz}-i\varepsilon_{yz})$ and $-id_{\mathrm{v}}\varepsilon_{xy}$, where $d_{\mathrm{v}}$ is a valence-band DP coefficient. In a homogeneous bulk system, where these bands are degenerate at the $\Gamma$ point of the Brillouin zone, by virtue of the degenerate perturbation theory, one would expect an energy splitting, hence a shift of the ground state
\begin{equation}\label{eq:shear-lh_deg}
    \Delta E_{\mathrm{sh,3D}} = d_{\mathrm{v}} \sqrt{|\varepsilon_{xy}|^2+|\varepsilon_{xz}|^2 + |\varepsilon_{yz}|^2}.
\end{equation}
In a low-dimensional system, the confinement (along with the built-in strain, if present) lifts the degeneracy, and the expected effect of the inter-subband strain-induced couplings follows from the second-order perturbation theory,
\begin{equation}\label{eq:shear-lh_split}
    \Delta E_{\mathrm{sh,low-D}} 
    \sim \frac{d^2_{\mathrm{v}}}{\Delta E_{\mathrm{lh}}} \left(|\varepsilon_{xy}|^2+|\varepsilon_{xz}|^2 + |\varepsilon_{yz}|^2 \right),
\end{equation}
where $\Delta E_{\mathrm{lh}}$ is a (rough) measure of the heavy-light hole subband splitting in a given system, which is typically on the order of tens of meV or more. This yields a small factor of $d_{\mathrm{v}}\varepsilon/\Delta E_{\mathrm{lh}}$, which may be expected to reduce the effect. Quantitative conclusions must be drawn from exact calculations to be presented in the subsequent sections.

\subsection{Piezoelectric coupling}
\label{sec:model-piezo}

Apart from the DP coupling, strain fields affect the charge state energies of the QD via strain-induced piezoelectric fields. We describe this effect in terms of the multipole expansion,
\begin{equation*}
\Delta E_{\mathrm{piezo}} = \Delta E_{\mathrm{piezo}}^{(1)} + \Delta E_{\mathrm{piezo}}^{(2)} + \ldots,
\end{equation*}
where the zeroth (monopole) term has been omitted in view of the global charge neutrality of the QD charge distribution. The dipole and quadrupole contributions to the electrostatic energy are
\begin{equation*}
\Delta E_{\mathrm{piezo}}^{(1)} = - \bm{p}\cdot \mathbcal{E},\quad
\Delta E_{\mathrm{piezo}}^{(2)} = -\frac{1}{12} \sum_{i,j} Q_{ij} 
\left( \frac{\partial \mathcal{E}_j}{\partial x_i} + \frac{\partial \mathcal{E}_i}{\partial x_j} \right),
\end{equation*}
with the dipole and quadrupole moments
\begin{align}\label{eq:dip_Quad}
\bm{p} & = \int d^3 r \bm{r} \rho_{\rm c}(\bm{r},\mathbcal{E}), \nonumber \\
Q_{ij} &= \int d^3 r \rho_{\rm c}(\bm{r},\mathbcal{E}) \left( 3 r_{i} r_{j} - r^2 \delta_{ij} \right),
\end{align}
where $\rho_{\rm c}(\bm{r})$ is the charge density. The field distributions associated with a given acoustic mode need to be found from the finite-element simulations described in Sec.~\ref{subsec:crystal}, while the multipole moments will be determined from exciton states obtained via combined \kp{}-CI calculations described in Sec.~\ref{subsec:kp_calc}. For the induced dipole moment, we use the linear polarization model, 
\begin{equation*}
\bm{p}(\mathbcal{E}) = \bm{p}_0 + \alpha \mathbcal{E},
\end{equation*}
with the polarizability tensor determined from the induced dipole moment, as found from \kp{}-CI calculations, at weak electric fields in the crystallographic directions. This leads to the total dipole component of the energy shift of the QD transition 
\begin{equation}
\label{eq:PiezoResponseDipole}
\Delta E_{\mathrm{piezo}}^{(1)}  = -\bm{p}_0\cdot \mathbcal{E} 
- \frac{1}{2}\mathbcal{E}^{\mathrm{T}} \alpha \mathbcal{E}.
\end{equation}
Similarly, one can include the built-in and induced quadrupole moment up to the first order in the field.

\section{Results}\label{sec:results}

\subsection{Phononic crystal and waveguide band structure}
\label{sec:band}

\begin{figure}[tb]
    \centering
	\includegraphics[width=\columnwidth]{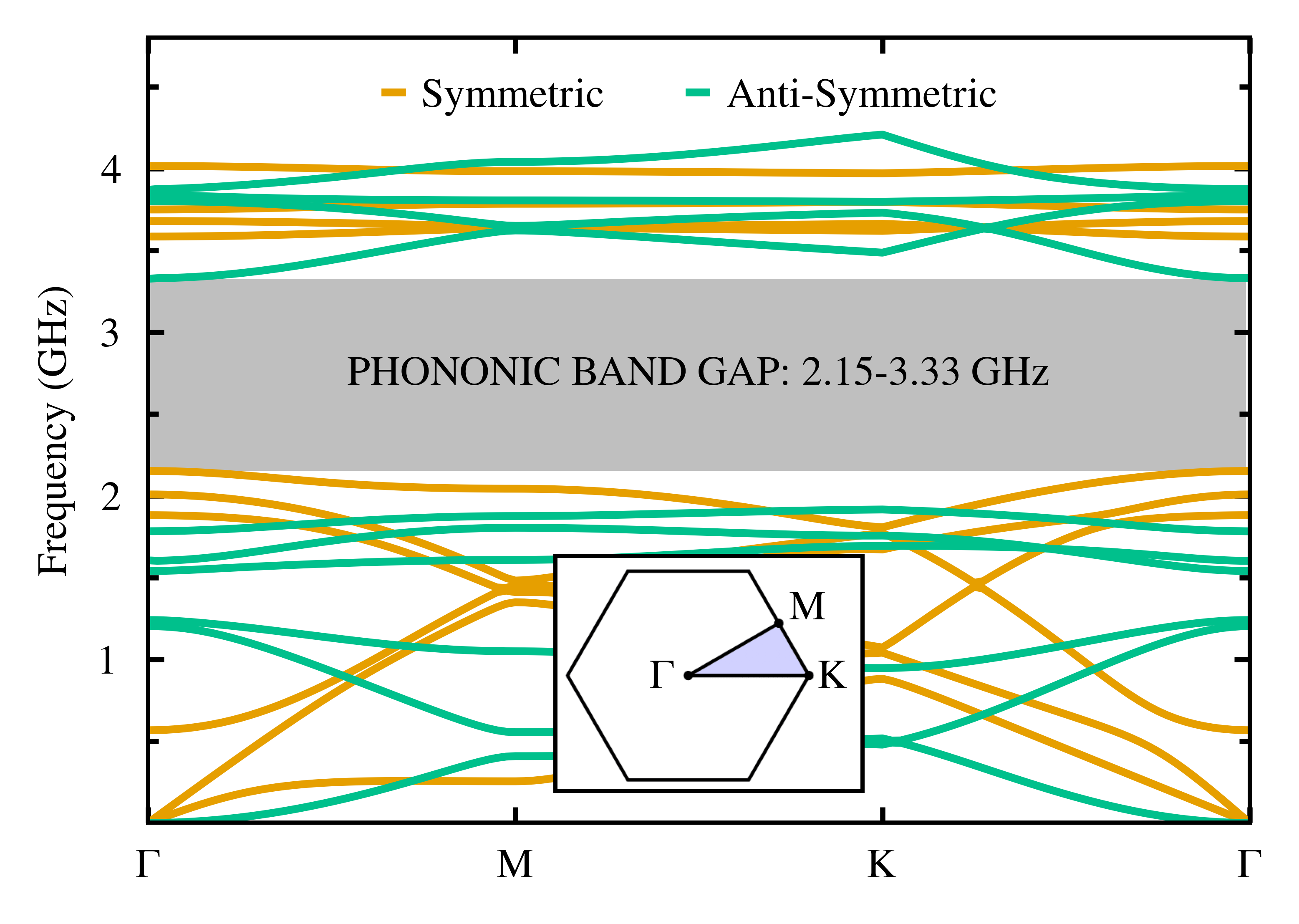}
	\caption{The phononic band structure of the snowflake PnC. Band diagrams were calculated for a GaAs structure characterized by parameters ($d$, $r$, $w$, $a$) = ($220$~nm, $0.44a$, $0.19a$, $760$~nm). Under these parameters, a significant phononic band gap (denoted by the gray shaded area) is present for both types of modes. Orange and green curves indicate modes that are symmetric and antisymmetric, respectively, with respect to the $xy$-plane.}
    \label{fig:fig3-band-structure}
\end{figure}

We start with optimizing the PnC geometry in order to provide a band structure featuring a broad phononic band gap in the GHz frequency range of interest, which will support the desired waveguide modes. A comprehensive numerical investigation allowed us to determine the geometric parameters as $a = 760$~nm, $r=0.44a$, $w = 0.19a$, and $d=220$~nm. These geometry parameters will be kept fixed throughout the following study. The band structure for such a crystal along the high-symmetry directions is shown in Fig.~\ref{fig:fig3-band-structure}. The plot displays the ten lowest symmetric modes (orange) and the ten lowest antisymmetric modes (green), with the symmetry defined with respect to the horizontal plane. A complete mechanical gap emerges between the sixth symmetric band and the seventh antisymmetric band, spanning from 2.15~GHz to 3.33~GHz, signifying a gap-to-midgap ratio of 43\%.

\begin{figure}[tb]
    \centering
    \includegraphics[width=\linewidth]{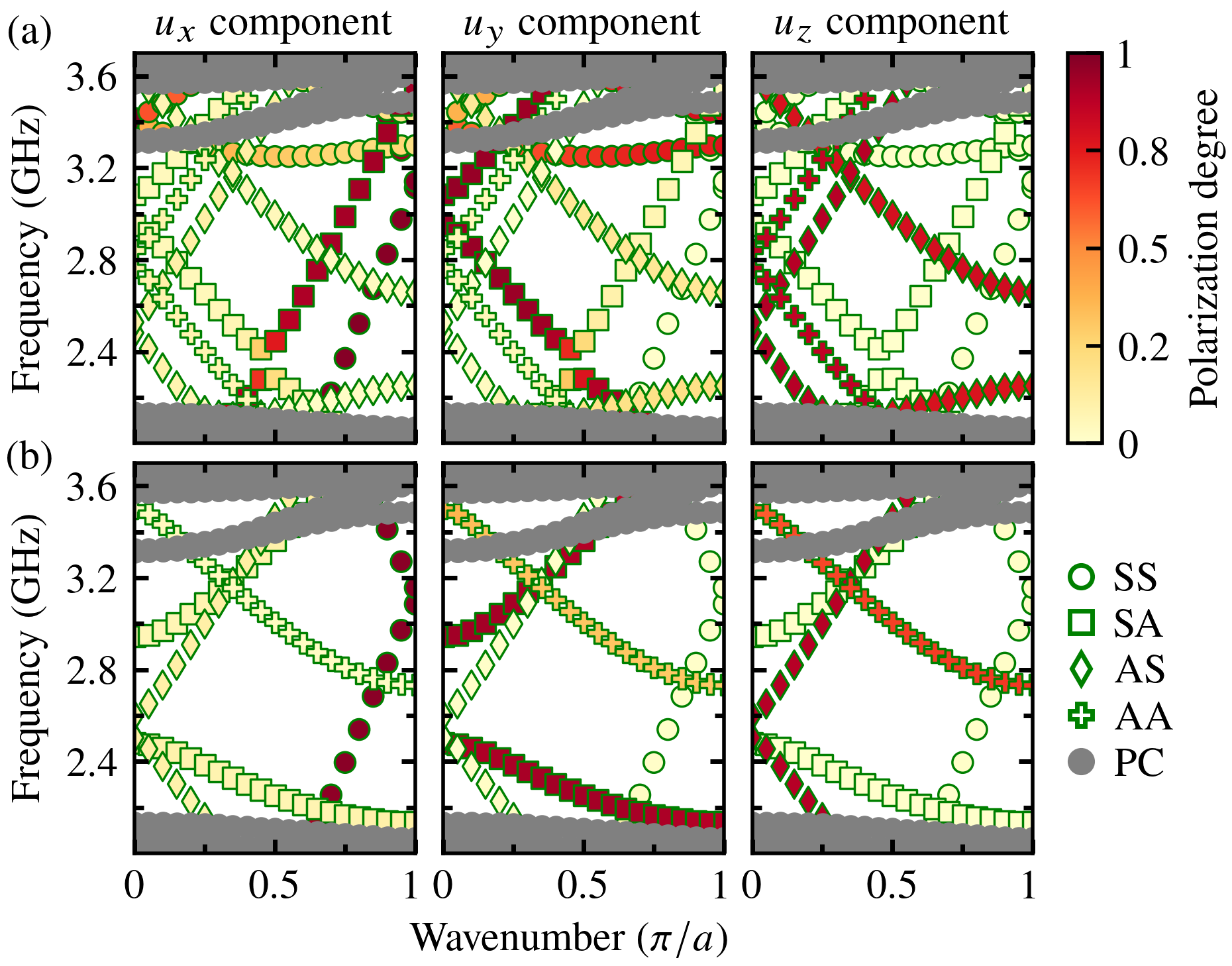}
    \caption{Dispersion curves of guided modes in a W1m waveguide for $\Delta = 0.3a$ (a) and $0.6a$ (b). Point shapes indicate the symmetry of the mode, as defined on the right. The color scale shows the contribution of the respective displacement components, $P_l$, averaged over the volume of a supercell, according to Eq.~\eqref{eq:Pl}.}
    \label{fig:fig4-skladoweCombined}
\end{figure}

As a next step, we determine the band structure of the waveguide. 
Figure~\ref{fig:fig4-skladoweCombined} shows the dispersion curves of guided modes for a W1m waveguide [see Fig.~\ref{fig:fig1-crystal-waveguide}(c)] at two values of $\Delta$. The dispersion relations are shown in three panels, each highlighting the relative contribution of one Cartesian component of the displacement [Eq.~\eqref{eq:Pl}] via the color scale (see the panel labels). The symmetry of the modes is indicated by symbols. Delocalized PnC modes are denoted by gray dots.

A relatively wide waveguide supports a large number of guided modes, which complicates selective excitation of the desired mode [Fig.~\ref{fig:fig4-skladoweCombined}(a)]. Narrowing the waveguide reduces the number of guided modes [Fig.~\ref{fig:fig4-skladoweCombined}(b)]. We therefore choose $\Delta = 0.6a$ for the subsequent analysis. 

Among the parameters relevant to the band structure, GaAs and Al$_{0.4}$Ga$_{0.6}$As differ significantly only in mass density (see Tab.~\ref{tab:params} in the Appendix~\ref{sec:mat-params}). The lower density of Al$_{0.4}$Ga$_{0.6}$As shifts its bands to higher frequencies. Since the two materials are otherwise practically identical in their band structure and modal properties, we present here only the results for the GaAs structure. For completeness, Appendix~\ref{sec:AlGaAs} demonstrates that an identically designed Al$_{0.4}$Ga$_{0.6}$As waveguide exhibits the same band structure and polarization properties, uniformly shifted to higher frequencies.

\subsection{QD susceptibility to strain and piezoelectric field}
\label{sec:QDsuscept}

The charge system confined in a QD responds to the strain associated with the mechanical wave directly, i.e., via conduction- and valence-band DPs, as well as via piezoelectric fields. Although this response can feature various subtle characteristics, here we are interested only in the energy shift of the fundamental bright excitonic transition. In this section, we characterize the QD response to various strain and piezoelectric field components by studying their effect on the fundamental transition energy in general. Based on this general knowledge, in Sec.~\ref{sec:QDResponse}, we will separately discuss the QD coupling to various strain components and electric fields in the waveguide. Finally, a combination of all effects together will be addressed in Sec.~\ref{sec:TotalResponse}.

\begin{figure}[tb]
	\includegraphics[width=\columnwidth]{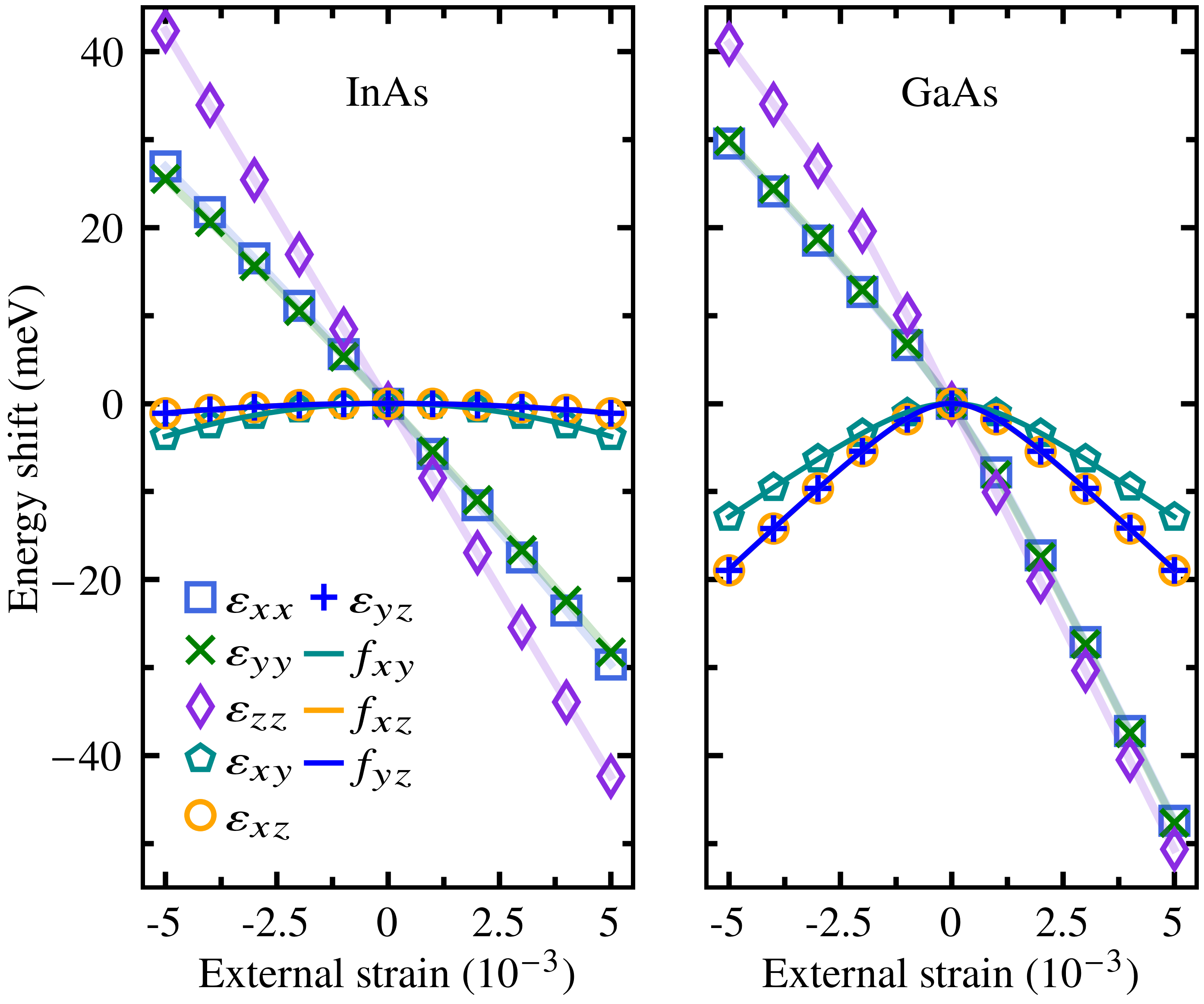}
	\caption{Modulation of the excitonic transition for the InAs (left) and GaAs (right) QD as a function of different components of external strain. Dark lines represent fitting functions $f_{ij}^{(\mathrm{InAs})}$ and $f_{ij}^{(\mathrm{GaAs})}$. Pale lines are added as a guide to the eye.} 
    \label{fig:fig5-Ex-both-InAs-GaAs_bR}
\end{figure}

\begin{table}[tb]
\begin{ruledtabular}
\begin{tabular}{lcccc}
 & \multicolumn{1}{c}{InAs} & \multicolumn{3}{c}{GaAs} \\
\colrule
 & \multicolumn{3}{c}{Diagonal components} \\
\colrule
$ii$ &
$A$ (eV) &
 & $A$ (eV) &  \\
\colrule
$xx$ & $-5.6612$ &  & $-7.4332$ &  \\
$yy$ & $-5.3769$ &  & $-7.5569$ &  \\
$zz$ & $-8.4755$ &  & $-9.9746$ &  \\
\colrule
 & \multicolumn{3}{c}{Shear components} \\
\colrule
$ij$ &
$B$ (eV) &
$\varepsilon_0$ & $B$ (eV) & $c$ (meV) \\
\colrule
$xy$ & $-150.27$ & $2.346\times10^{-6}$ & 1091.83 & 5.777 \\
$xz$ & $-44.29$  & $-1.768\times10^{-6}$ & 1972.36 & 5.873 \\
$yz$ & $-44.17$  & $1.191\times10^{-6}$ & 1946.63 & 5.990 \\
\end{tabular}
\end{ruledtabular}
\caption{\label{tab:fitting-strain}
Fitting parameters describing the excitonic energy shift induced by diagonal
(top) and shear (bottom) strain components.
Shear components are fitted using Eqs.~\eqref{eq:fij-InAs} and~\eqref{eq:fij-GaAs},
while diagonal components are described by a linear dependence [Eq.~\eqref{eq:diagElements}].}
\end{table}

We calculate the DP effect directly using our \kp{}-CI simulations for a QD subject to a single-component strain. The results for the two types of QDs are presented in Fig.~\ref{fig:fig5-Ex-both-InAs-GaAs_bR}. Both QDs show a strong response to the individual diagonal components of the strain tensor. Within our range of strain, this dependence is fitted with a linear function
\begin{equation}\label{eq:diagElements}
g_{ii}(\varepsilon) = A\,\varepsilon,
\end{equation}
where $i=x,y,z$. The fit is performed using the five central strain points and the resulting fitting parameters are listed in Table~\ref{tab:fitting-strain}. The diagonal response can also be decomposed into a component proportional to the volumetric strain [see Eq.~\eqref{eq:Evol}], with the excitonic DP constant $D_{\rm vol}\approx -6.5$~eV and $-8.3$~eV for InAs and GaAs QD, respectively, and a biaxial component, $\Delta E_{\rm biax} = D_{\rm biax}(2\varepsilon_{zz}-\varepsilon_{xx}-\varepsilon_{yy})$, with $D_{\rm biax}\approx -1.0$~eV and $-0.8$~eV for InAs and GaAs QD, respectively. In the case of the GaAs QD, a non-linear bending of the response curves appears already at moderate negative strains. The single-particle strain susceptibility, presented in Appendix~\ref{sec:SPB}, shows that the biaxial contribution, as well as the deviations from linearity, come from the valence band as expected.

The response to shear strain components differs considerably for the two structures, which is a result of their very different confinement geometry. The InAs dot displays only a weak response to the \(\varepsilon_{xy}\) component, while the effect of the other two shear components is negligible. In contrast, the GaAs QD responds stronger to all shear components. As shown in Appendix~\ref{sec:SPB}, the shear-strain response originates from the valence band, as expected.
The effect of shear strain shows a clear parabolic behavior near the apex, indicating the heavy-light hole splitting in line with Eq.~\eqref{eq:shear-lh_split}. For the GaAs QD, this is limited to weak strains with clear linear asymptotics, consistent with the weaker vertical confinement and lack of strain, and hence a reduced heavy-light hole splitting in such structures. Consequently, we fit the \kp{}-CI results with the parabolic function 
\begin{equation}\label{eq:fij-InAs}
f_{ij}^{(\mathrm{InAs})}(\varepsilon) = B\varepsilon^2
\end{equation}
and the two-level-resonance function 
\begin{equation}\label{eq:fij-GaAs}
f_{ij}^{(\mathrm{GaAs})}(\varepsilon) = c-\sqrt{c^2 + 2Bc\varepsilon (\varepsilon-2\varepsilon_0)},
\end{equation}
$ij = xy,xz,yz$, for the InAs and GaAs QDs, respectively (solid lines in Fig.~\ref{fig:fig5-Ex-both-InAs-GaAs_bR}). Here, we allow for a shifted position of the apex and at the same time require that $f_{ij}^{(\mathrm{InAs})}(0) = f_{ij}^{(\mathrm{GaAs})}(0) = 0$. The parameters resulting from least-squares fitting are listed in Tab.~\ref{tab:fitting-strain}.

\begin{table}[tb]
\begin{ruledtabular}
\begin{tabular}{lcc}
\textrm{Parameter} & \textrm{InAs} & \textrm{GaAs} \\
\colrule
$p_x^{(0)}$ (nm$\cdot e$) & $0$  & $-1.051\times 10^{-3}$  \\
$p_y^{(0)}$ (nm$\cdot e$) & $0$  & $7.439 \times 10^{-3}$   \\
$p_z^{(0)}$ (nm$\cdot e$) & $-0.1632$  & $-6.4904 \times 10^{-2}$  \\
\colrule
$\alpha_{xx}$  ($e\cdot$nm$^2/$mV) & 2.586  & 4.652 \\
$\alpha_{yy}$  ($e\cdot$nm$^2/$mV) &  2.858 & 4.448 \\
$\alpha_{zz}$  ($e\cdot$nm$^2/$mV) & 0.0629 & 0.3155 \\ 
\end{tabular}
\end{ruledtabular}
\caption{\label{tab:dipoleAndquadrupol}%
Zero-field dipole moments and the diagonal elements of the dipolar polarizability tensor for InAs and GaAs QDs.}
\end{table}

\begin{figure}[tb]
	\includegraphics[width=\columnwidth]{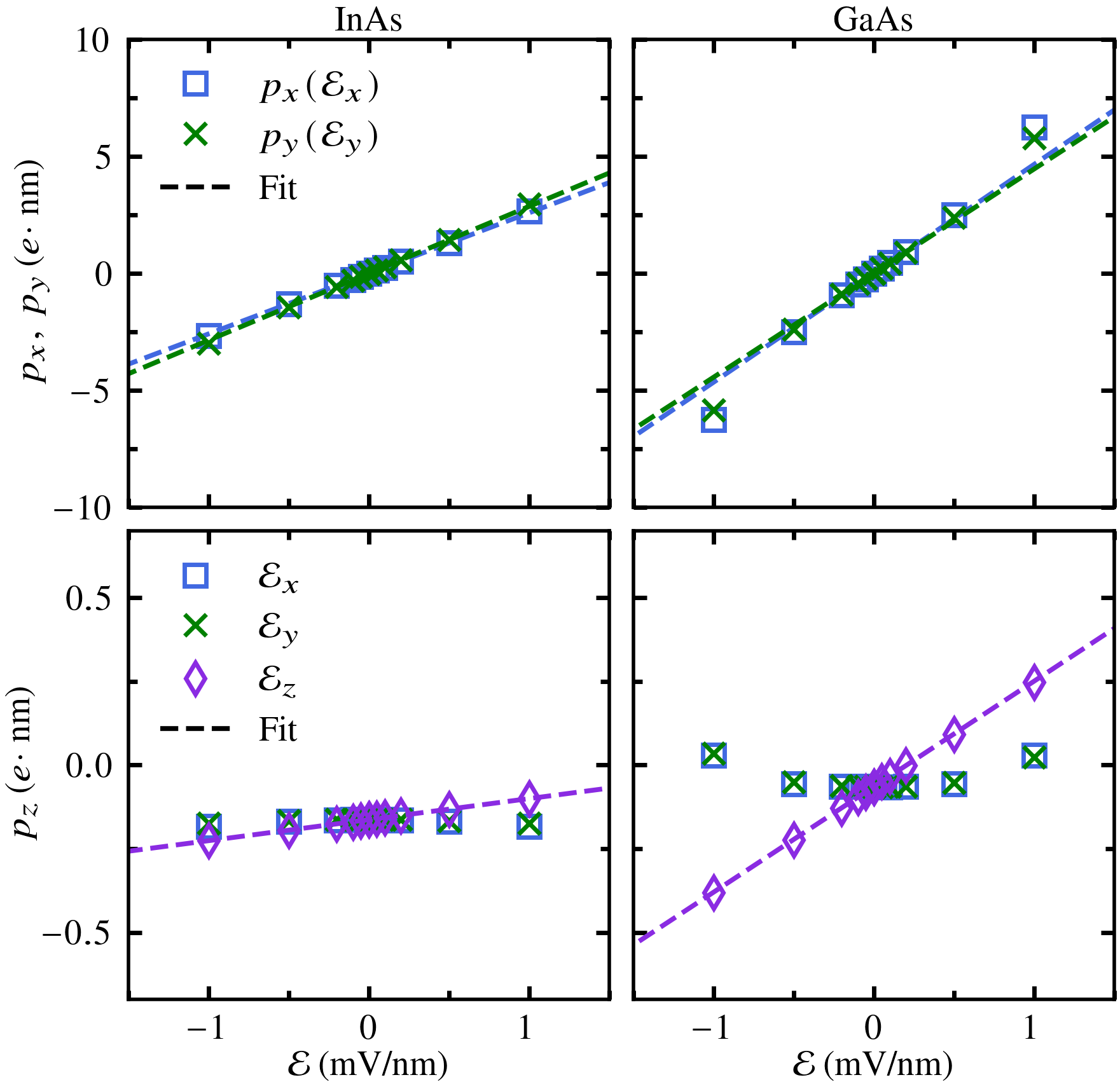}
	\caption{Electric-field dependence of the excitonic dipole moment for the dominant components in the InAs (left) and GaAs (right) QDs. Different markers represent three distinct diagonal responses to the dominant components of the electric field. Dashed lines show the linear fits based on 7 central data points.} 
    \label{fig:fig6-dipole_response}
\end{figure}

In view of the small size of the QD, the interaction with electric fields is treated via multipole expansion, as discussed in Sec.~\ref{sec:model-piezo}. Again, we use \kp{}-CI calculations to determine the electric-field dependence of the excitonic dipole moment, as shown for the leading components in Fig.~\ref{fig:fig6-dipole_response}. From this, we deduce the zero-field dipole moments for the two QDs, as well as the diagonal elements of the polarizability tensor, which we list in Tab.~\ref{tab:dipoleAndquadrupol} in the basis $x,y,z$, aligned with the waveguide structure. The values were extracted by a linear fit to the computed dependence $p_j(\mathcal{E}_j)$, $j=x,y,z$, based on 7 central data points (shown by dashed lines in Fig.~\ref{fig:fig6-dipole_response}). Comparison to the exact data in Fig.~\ref{fig:fig6-dipole_response} shows that this fit remains quite accurate for fields up to 1~mV/nm. The zero-field dipole for the InAs QD is vertical up to numerical precision, in agreement with the flat elliptical geometry of our QD model. For the GaAs QD, the more isotropic confinement in combination with random shape fluctuations, faithfully mapped onto the \kp{} model, leads to a substantial tilt along an axis that does not coincide with any principal crystal direction. The $\alpha_{xy}$ elements of the polarizability tensor are at least two orders of magnitude smaller for the GaAs QD and within the numerical noise for the InAs QD and will be neglected in our analysis.
The data sheets with all the numerical results underlying this analysis are available as Supplementary Material to this paper~\cite{sm2026}. The dependence of the $z$ component of the dipole moment on the in-plane electric field is quadratic (see the lower panel of Fig.~\ref{fig:fig6-dipole_response}) and adds a correction on the level of a fraction of the built-in dipole for the fields considered here. We will neglect this quadratic correction, which might lead to a small cubic correction to the energy shift if two different components of the electric field are sufficiently strong. 

The spatial variation of the electric field takes place on the length scales of the supercell size, membrane thickness, or waveguide width, all of them much larger than the QD size. Therefore, we expect the multipole expansion to be convergent and the next-order, quadrupole terms to yield only small corrections. In Appendix~\ref{appendix:quadrupole} we show that this is indeed the case. Therefore, we restrict the piezoelectric couplings to the dipole term in our discussion.

\subsection{QD response to strain and strain-related fields in the waveguide}
\label{sec:QDResponse}

\begin{table}[tb]
\begin{ruledtabular}
\begin{tabular}{lccc}
\textrm{Mode} & symmetry & $k$ & $f$ \\
\colrule
LS-like & SS & $0.8\pi/a$ & $2.54$~GHz \\
LA-like & AS & $0.15\pi/a$ & $2.30$~GHz \\
SH-like & SA & $0.15\pi/a$ & $3.00$~GHz \\
mixed LA-SH & AA & $0.8\pi/a$ & $2.78$~GHz 
\end{tabular}
\end{ruledtabular}
\caption{\label{tab:modes}%
Four representative vibration modes of the GaAs membrane and their corresponding frequencies.}
\end{table}

In this section, we study the QD response to the strain and piezoelectric fields associated with a guided wave. For the QD, we allow only spatial positions located at least $50$~nm from any air--membrane interface, to ensure that QD experiences a uniform local environment and maintains stable emission properties. This sets, in particular, the upper limit for the QD positions in the vertical direction, which we denote $z_{\mathrm{max}}$. For our analysis, we choose four modes that represent different symmetries, as shown in Tab.~\ref{tab:modes}. Since an individual mode usually does not cover the full extent of the Brillouin zone (see Fig.~\ref{fig:fig4-skladoweCombined}), we need to take these modes at two different $k$ points. We discuss the QD response separately for the three coupling channels that coexist in various proportions for each mode: DP response to volumetric strain, DP response to shear strain, and dipole response including polarizability. 

\begin{figure}[tb]
	\includegraphics[width=\columnwidth]{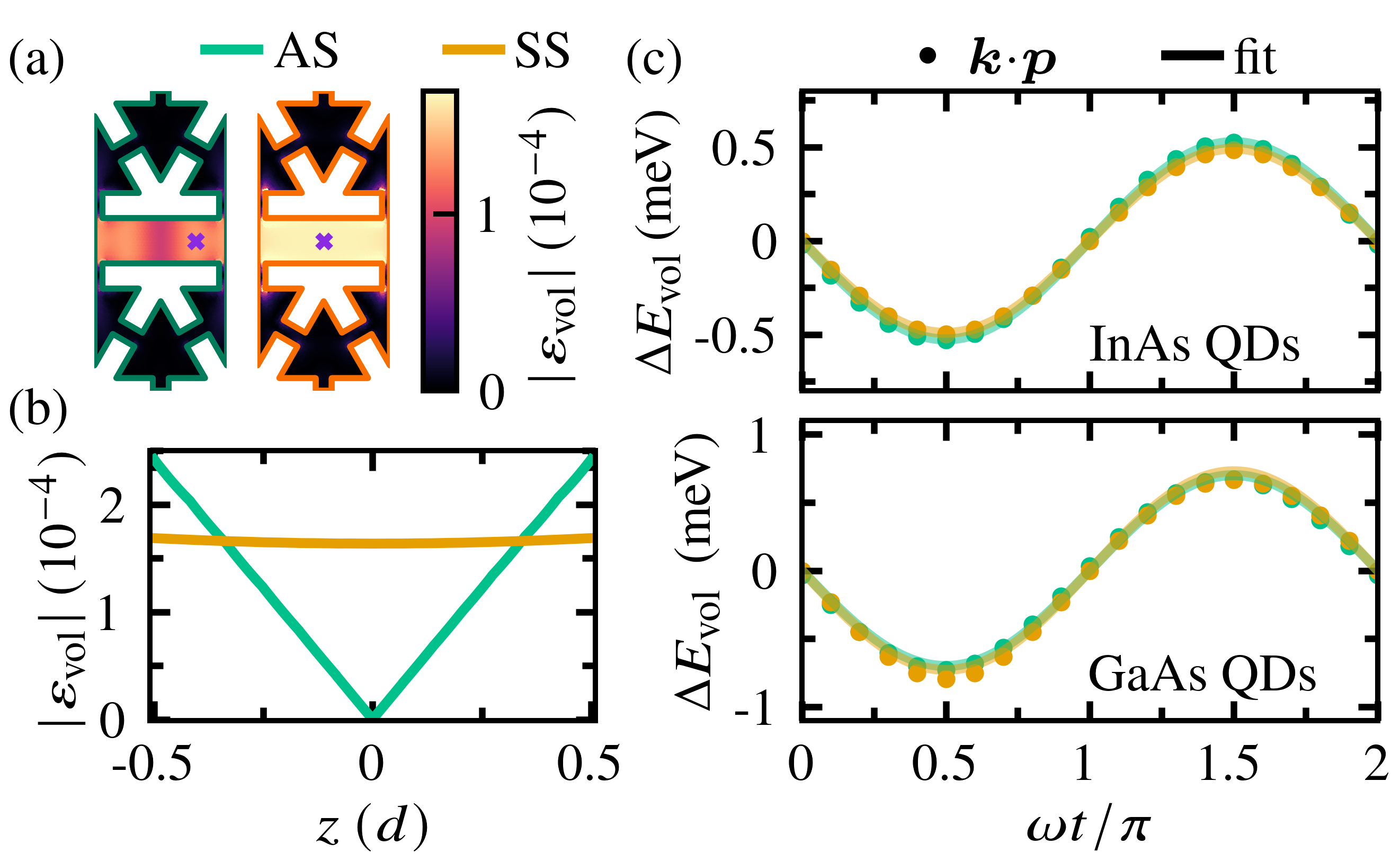}
	\caption{(a) Amplitude of the volumetric strain for the two modes as indicated by the color-coded border. Crosses mark the $xy$ position corresponding to the highest strain amplitude. (b) Dependence of the amplitude of volumetric strain on the vertical position for the selected $xy$ position. (c) Evolution of the strain-induced energy shift due to diagonal DP coupling over one period of oscillation. Points show \kp{}-CI results, lines show a simple harmonic fit as described in the text. Colors refer to the two modes.}\label{fig:fig7-wolumetryczne_AS_SS}
\end{figure}

\begin{table}[tb]
\begin{ruledtabular}
\begin{tabular}{lccccc}
 & \multicolumn{3}{c}{Point} & \multicolumn{2}{c}{$|\tilde{\varepsilon}_{\rm vol}|$} \\
\colrule
mode & $x$ & $y$ & $z$ & GaAs & AlGaAs \\
\colrule
AS & $0.262a$ & 0 & $z_{\mathrm{max}}$ & $1.334\times 10^{-4}$ & $1.301\times 10^{-4}$ \\
SS & 0 & 0 & $z_{\mathrm{max}}$ & $1.650\times 10^{-4}$ & $ 1.639\times 10^{-4}$ \\
\end{tabular}
\end{ruledtabular}
\caption{Points of the maximum volumetric strain amplitude and the value of this amplitude for two different membranes used in the simulations for diagonal strain elements.}
\label{tab:max-strain-diag}%
\end{table}

The spatial profiles of all strain components of the modes can be found in Appendix~\ref{appendix:ModeProfiles}. With our normalization of maximum displacement to 0.1~nm, the strain values reach $2$--$3 \times 10^{-4}$, with the highest volumetric strain amplitude occurring for the SS mode. These values are nearly an order of magnitude larger than those deduced for a similar membrane with InGaAs QDs, where the energy shift was $\pm 27~\mu$eV~\cite{carter2017sensing}. As seen in the Appendix~\ref{appendix:ModeProfiles}, the diagonal strain components that contribute to volumetric and axial strain are significant only for modes with SS (LS-like) and SA (LA-like) symmetry. For the SS mode, the strains are essentially homogeneous within the waveguide core and largely independent of the vertical position [see Fig.~\ref{fig:fig7-wolumetryczne_AS_SS}(b)]. In contrast, the asymmetry of the AS mode imposes a strain-free condition (with respect to diagonal strain components) on the membrane symmetry plane at $z=0$, leading to non-zero strains only away from this plane. Moreover, the AS mode exhibits a higher-order profile with a full oscillation across the supercell, implying a position-dependent response of the QD. In both cases, the $\varepsilon_{zz}$ component is approximately twice weaker than the dominant $\varepsilon_{xx}$, with negligible $\varepsilon_{yy}$, consistent with the dominant Lamb-like character of the mode. The resulting volumetric strain amplitudes at $z = z_{\mathrm{max}}$ are shown in Fig.~\ref{fig:fig7-wolumetryczne_AS_SS}(a). For further discussion, we assume that the QD is located at a point where the volumetric strain reaches its maximum (indicated by crosses in Fig.~\ref{fig:fig7-wolumetryczne_AS_SS}(a); the exact spatial components of this point are listed in Tab.~\ref{tab:max-strain-diag}). Given the nearly linear response of the QD transition energy to the diagonal strain components (see Fig.~\ref{fig:fig5-Ex-both-InAs-GaAs_bR}), a harmonic modulation of the transition energy is expected in response to a harmonic strain field, with amplitude determined by the strain amplitude and the DP constant $D$. This expectation is confirmed in Fig.~\ref{fig:fig7-wolumetryczne_AS_SS}(c), which shows the energy shift at several time steps calculated using the \kp{}-CI method, with the strain field computed from the instantaneous displacement [see Eq.~\eqref{eq:u}] using the amplitudes found from finite-element simulations. For comparison, we also show results using the simplified formula $\Delta E(t) = E_0 \sin(2\pi f t)$, with the amplitude $E_0$ equal to $E_0^{\mathrm{AS}} = -0.527$~meV and $E_0^{\mathrm{SS}} = -0.491$~meV for the InAs QD, and $E_0^{\mathrm{AS}} = -0.700$~meV and $E_0^{\mathrm{SS}} = -0.729$~meV for the GaAs QD.
\begin{table}[tb]
\begin{ruledtabular}
\begin{tabular}{llccccc}
 & & \multicolumn{3}{c}{Point} & \multicolumn{2}{c}{Shear strain amplitude} \\
\colrule
mode & comp. & $x$ & $y$ & $z$ & GaAs & AlGaAs \\
\colrule
AS & $\varepsilon_{xz}$ & 0 & 0 & 0 & $1.634\times 10^{-4}$ & $1.638\times 10^{-4}$\\
SA & $\varepsilon_{xy}$ & $0.189a$ & 0 & $z_{\mathrm{max}}$ & $2.811\times 10^{-4}$ & $2.833\times 10^{-4}$\\
AA & $\varepsilon_{xy}$ & 0 & 0 & $z_{\mathrm{max}}$ & $1.787\times 10^{-4}$ & $1.742\times 10^{-4}$\\
\end{tabular}
\end{ruledtabular}
\caption{Dominant strain components and points of the maximum shear strain amplitude and the value of this amplitude for two different membranes used in the simulations.}
\label{tab:max-strain-shear}%
\end{table}

\begin{figure}[tb]
	\includegraphics[width=\columnwidth]{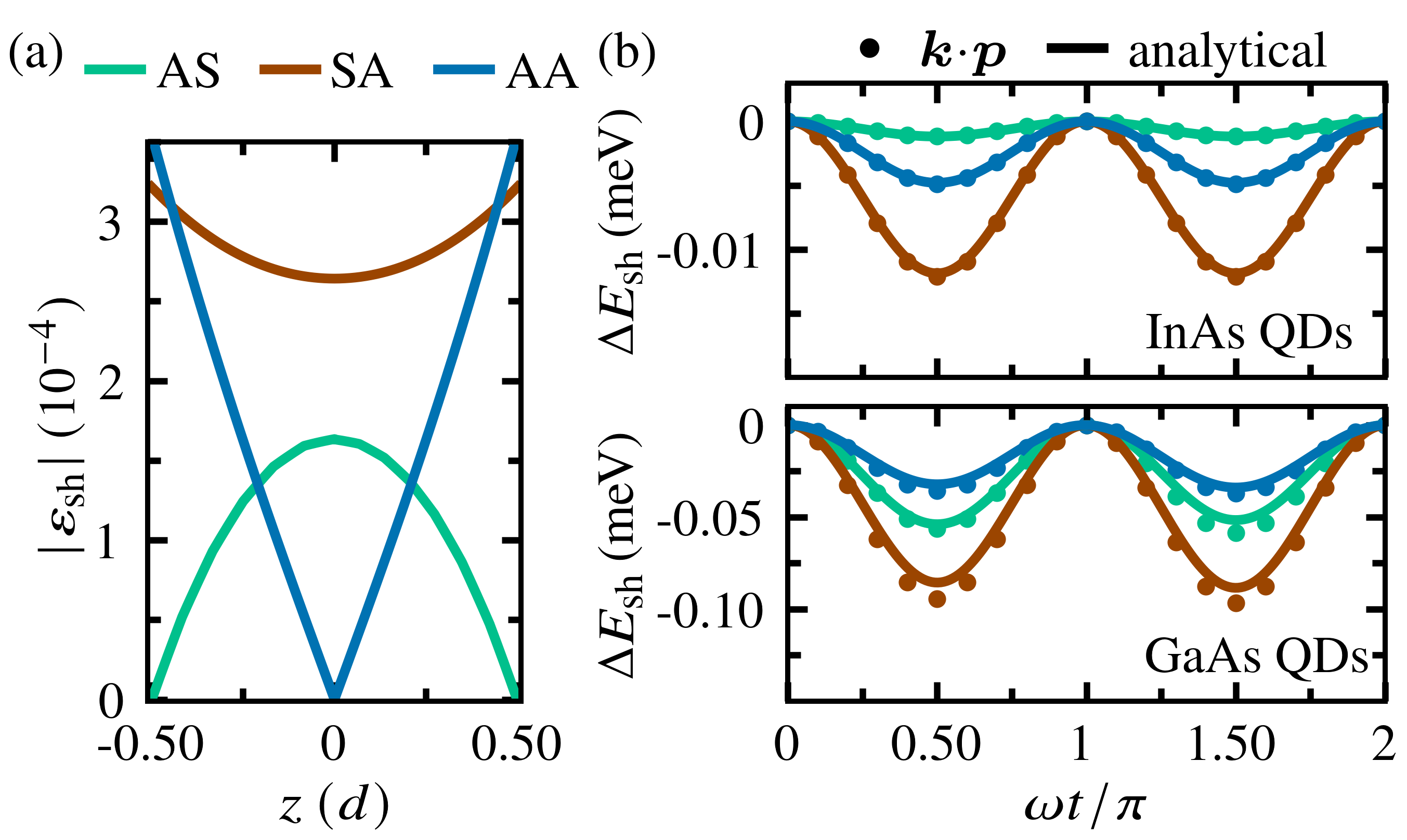}
	\caption{(a) Dependence of the amplitude of the dominant shear–strain component, $\varepsilon_{\mathrm{sh}}$, on the vertical coordinate $z$ for the three modes indicated. Each curve is taken at the in-plane point of maximum shear amplitude, and the corresponding $(x,y)$ coordinates and shear components are listed in Table~\ref{tab:max-strain-shear}. (b) Evolution of the strain-induced energy shift due to off-diagonal DP coupling over one period of oscillation. Points show \kp{}-CI results, lines show the results obtained by the analytical function as described in the text. Colors refer to the three modes. }\label{fig:fig8-scinajace_AS_SA_AA}
\end{figure}

Shear strain reaches significant values for three elastic modes: AS, SA, and AA, each dominated by different shear strain components and characterized by distinct wave profiles along the waveguide, as shown in the strain maps in Appendix~\ref{appendix:ModeProfiles}, and different strain distributions in the vertical direction, presented in Fig.~\ref{fig:fig8-scinajace_AS_SA_AA}(a). Specifically, the AS mode exhibits a dominant $\varepsilon_{xz}$ component, while the SA and AA modes are dominated by $\varepsilon_{xy}$. In the SA mode, which is symmetric with respect to the horizontal mirror plane $\sigma_z$, the $\varepsilon_{xy}$ component is nearly uniform across the membrane core, similar to the volumetric strain observed in the SS mode. In contrast, the AS and AA modes show more complex, spatially varying shear strain distributions, leading to a strongly position-dependent QD response. By combining the strain distribution maps with the QD strain susceptibilities from Fig.~\ref{fig:fig5-Ex-both-InAs-GaAs_bR}, it becomes clear that for the AS mode, the effect of shear strain is relatively weak away from the midplane of the membrane and negligible compared to the volumetric strain. However, at the midplane (where the volumetric strain vanishes due to symmetry), shear strain becomes the dominant contribution. For the SA and AA modes, the shear component is the primary driver of the QD response throughout the relevant regions. For each mode, we selected the spatial points at which the dominant shear strain component reaches its maximum; these locations are listed in Table~\ref{tab:max-strain-shear}. Importantly, the QD response to shear strain differs qualitatively from its response to diagonal components: due to the quadratic dependence of the energy shift on shear strain, the QD transition energy exhibits a dominant second harmonic response (i.e., at $2\omega$). This frequency-doubling behavior is clearly seen in Fig.~\ref{fig:fig8-scinajace_AS_SA_AA}(b) (points), where we show the energy modulation computed, as before, using the \kp{}-CI method point-by-point, but now including only the shear strain contribution. Since the shear strain is dominated by a single component in all these cases, we can also substitute Eq.~\eqref{eq:strain-evol} into the susceptibility functions $f_{ij}^{(\mathrm{InAs})}$ and $f_{ij}^{(\mathrm{GaAs})}$ from  Eq.~\eqref{eq:fij-InAs} and Eq.~\eqref{eq:fij-GaAs}, with $ij$ representing the dominant shear strain component, to compute the energy shifts analytically. These results are also shown in Fig.~\ref{fig:fig8-scinajace_AS_SA_AA}, reproducing the exact numerical values very well.

\begin{table}[tb]
\begin{ruledtabular}
\begin{tabular}{llccccc}
 & & \multicolumn{3}{c}{Point} & \multicolumn{2}{c}{Elec. field ampl. (mV/nm)} \\
\colrule
mode & comp. & $x$ & $y$ & $z$ & GaAs & AlGaAs\\
\colrule
SS & $\mathcal{E}_{z}$ & 0 & 0 & $z_{\mathrm{max}}$ & $0.590$ & $0.466$\\
AS & $\mathcal{E}_{x}$/\,$\mathcal{E}_{z}$  & 0 & 0 & $z_{\mathrm{max}}$ & $0.356/\, 0.245$ & $0.276/\, 0.188$\\
SA & $\mathcal{E}_{y}$ & $a/2$ & 0 & $z_{\mathrm{max}}$ & $0.207$ & $0.172$\\
AA & $\mathcal{E}_{y}$ & $a/2$ & 0 & $z_{\mathrm{max}}$ & $0.189$ & $0.142$\\
\end{tabular}
\end{ruledtabular}
\caption{
Dominant electric field components and points of the maximum amplitude and the value of this amplitude for two different membranes used in the simulations.}
\label{tab:max-Efield}
\end{table}

For the discussion of piezoelectric coupling, we present an overview of the electric fields associated with the four modes in Appendix~\ref{appendix:ModeProfiles}. The electric fields are approximately 30\% stronger in the GaAs membrane, reaching a maximum of 0.590~mV/nm for the SS mode. 
As discussed in Sec.~\ref{sec:model-piezo} [see Eq.~\eqref{eq:PiezoResponseDipole}], the QD responds to the electric field through two mechanisms: a built-in dipole (with essentially only a $z$-component in both QDs), which couples linearly to the out-of-plane field $\mathcal{E}_z$, and an induced dipole which, to the leading order, leads to a quadratic response to the in-plane field components, mostly $\mathcal{E}_x$ and $\mathcal{E}_y$. As the electromechanical problem is fully linear, the electric field amplitudes scale linearly with the mechanical normalization. As discussed in Sec.~\ref{sec:QDsuscept}, the $z$ component of the dipole is dominated by its built-in, zero-field value. On the contrary, the in-plane components have very small zero-field values and show linear polarizability behavior that dominates the dipole response already at weak fields. As a result, $\mathcal{E}_z$ leads to a harmonic modulation of the transition energy, while $\mathcal{E}_x$ and $\mathcal{E}_y$ produce an asymmetric energy response and frequency doubling, analogous to the case of shear strain. 

The electric field distributions differ considerably between the modes. The SS mode exhibits a dominant $\mathcal{E}_x$ component that is nearly uniform in both the lateral and vertical directions, much like the volumetric strain for this mode. The AS mode exhibits both $\mathcal{E}_x$ and $\mathcal{E}_z$ components away from the midplane $z=0$, with $\mathcal{E}_z$ vanishing at the midplane due to symmetry. For the remaining two modes, SA and AA, the dominant field component in the region of interest is $\mathcal{E}_y$, which vanishes at $z=0$ for the SA mode but remains nearly constant with vertical position in the AA mode. The points where these dominant electric field components reach their maximum values were selected for further analysis and are listed in Table~\ref{tab:max-Efield}. In fact, the maximum of the $\mathcal{E}_x$ component for the AS mode is found at $x = 0.42a$ and reaches $0.364$~mV/nm, only marginally exceeding the value at $x = 0$ ($0.356$~mV/nm). Since the second relevant component, $\mathcal{E}_z$, exhibits substantially higher amplitudes at $x = 0$ than at $x = 0.42a$, this position was selected to evaluate the combined response of both components.

\begin{figure}[tb]
	\includegraphics[width=\columnwidth]{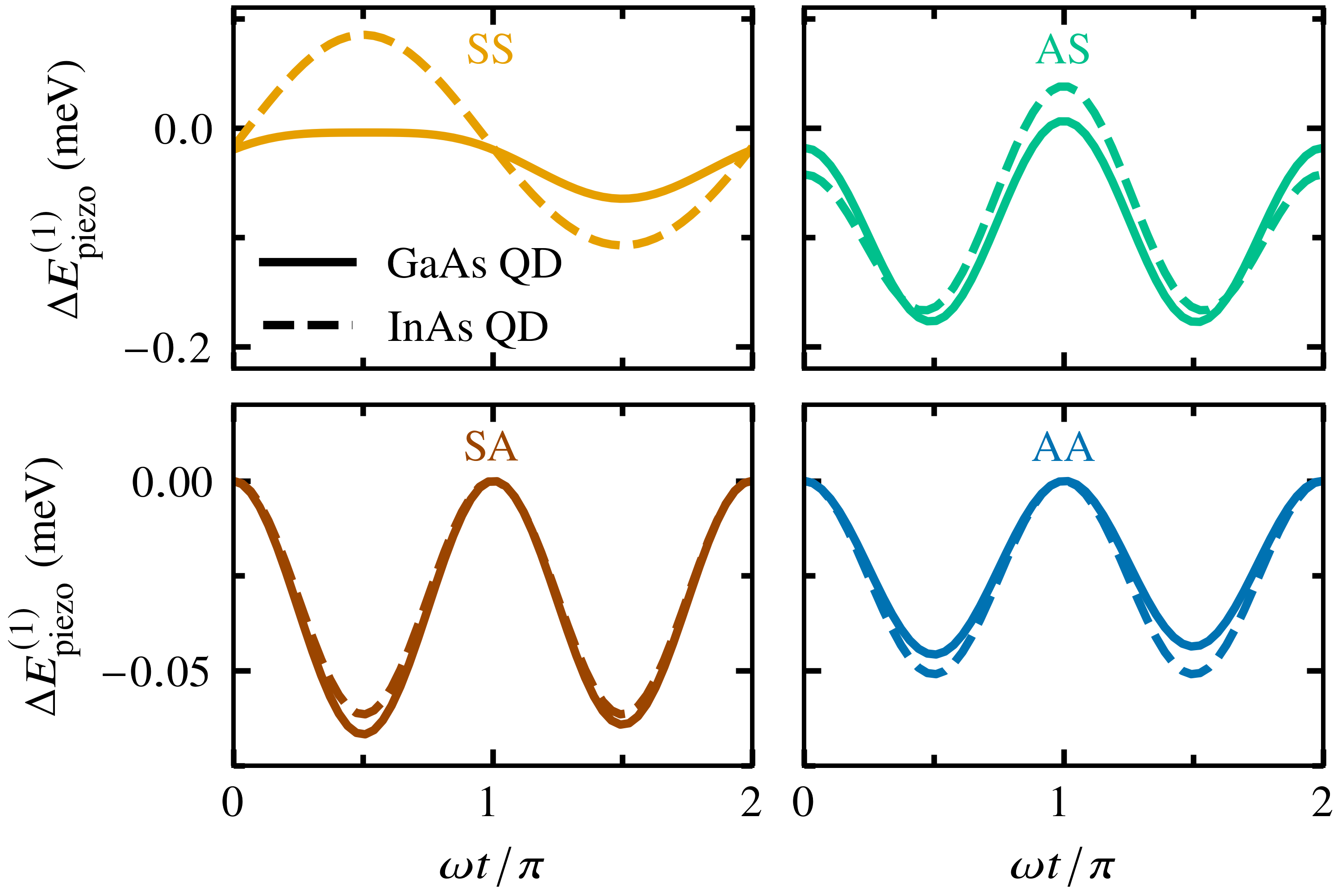}
	\caption{Time evolution of the piezoelectric contribution to the excitonic energy shift over one oscillation period for the four selected modes. Solid and dashed lines correspond to GaAs and InAs QDs, respectively. Colors distinguish the modes.}\label{fig:fig9-piezoResponse}
\end{figure}

Combining the field magnitudes with the dipole and polarizability tensor values in Tab.~\ref{tab:dipoleAndquadrupol}, we find that the response of the QD differs between the SS mode and the three remaining modes due to the nature of the dominant electric field components. A quantitative analysis is presented in Fig.~\ref{fig:fig9-piezoResponse}, where we plot the time dependence of the piezoelectric-field-induced dipole component to the energy shift, calculated from Eq.~\eqref{eq:PiezoResponseDipole} using the electric fields from the finite-element simulations [Eq.~\eqref{eq:Ei}]. In the SS mode, the out-of-plane ($z$) component dominates, leading to a predominantly harmonic modulation of the QD transition energy, as the coupling is governed by the built-in dipole. This results in maximum energy shifts of approximately 65~$\mu$eV (107~$\mu$eV) for GaAs (InAs) QDs, as shown in the upper left panel of Fig.~\ref{fig:fig9-piezoResponse}. In contrast, the remaining three modes are dominated by in-plane field components, which strongly polarize the QD. This leads to a nonlinear, quadratic response and, consequently, to frequency doubling of the energy modulation. The strongest in-plane field is found in the AS mode, where the electric field reaches up to 0.36~mV/nm, resulting in energy shifts of about 0.18~meV (0.16~meV) for GaAs (InAs) QDs. 

\subsection{Total QD response}
\label{sec:TotalResponse}

Finally, we evaluate the energy modulation of the QD for all four acoustic modes at spatial positions that maximize the dominant coupling mechanism for each mode. For the SS and AS modes, these are the points where the volumetric strain reaches its maximum, since the dominant coupling mechanism is the DP (Tab.~\ref{tab:max-strain-diag}). For the AA mode, the piezoelectric coupling is strongest, therefore the analysis is carried out at the location of the maximal relevant electric field component (Tab.~\ref{tab:max-Efield}). In contrast, for the SA mode, the dominant coupling mechanism is via the shear strain channel, which exceeds both the deformation potential and piezoelectric contributions. Consequently, for this mode, the modulation is evaluated at the position where the shear strain is maximized (Tab.~\ref{tab:max-strain-shear}).

We focus on a GaAs QD. In this material, strain-mediated couplings for both volumetric and shear components are enhanced relative to an InAs QD. This enhancement originates from the stronger intrinsic strain susceptibility of GaAs. In contrast, the piezoelectric coupling strengths are comparable between the two material systems.

\begin{figure}[tb]
	\includegraphics[width=\columnwidth]{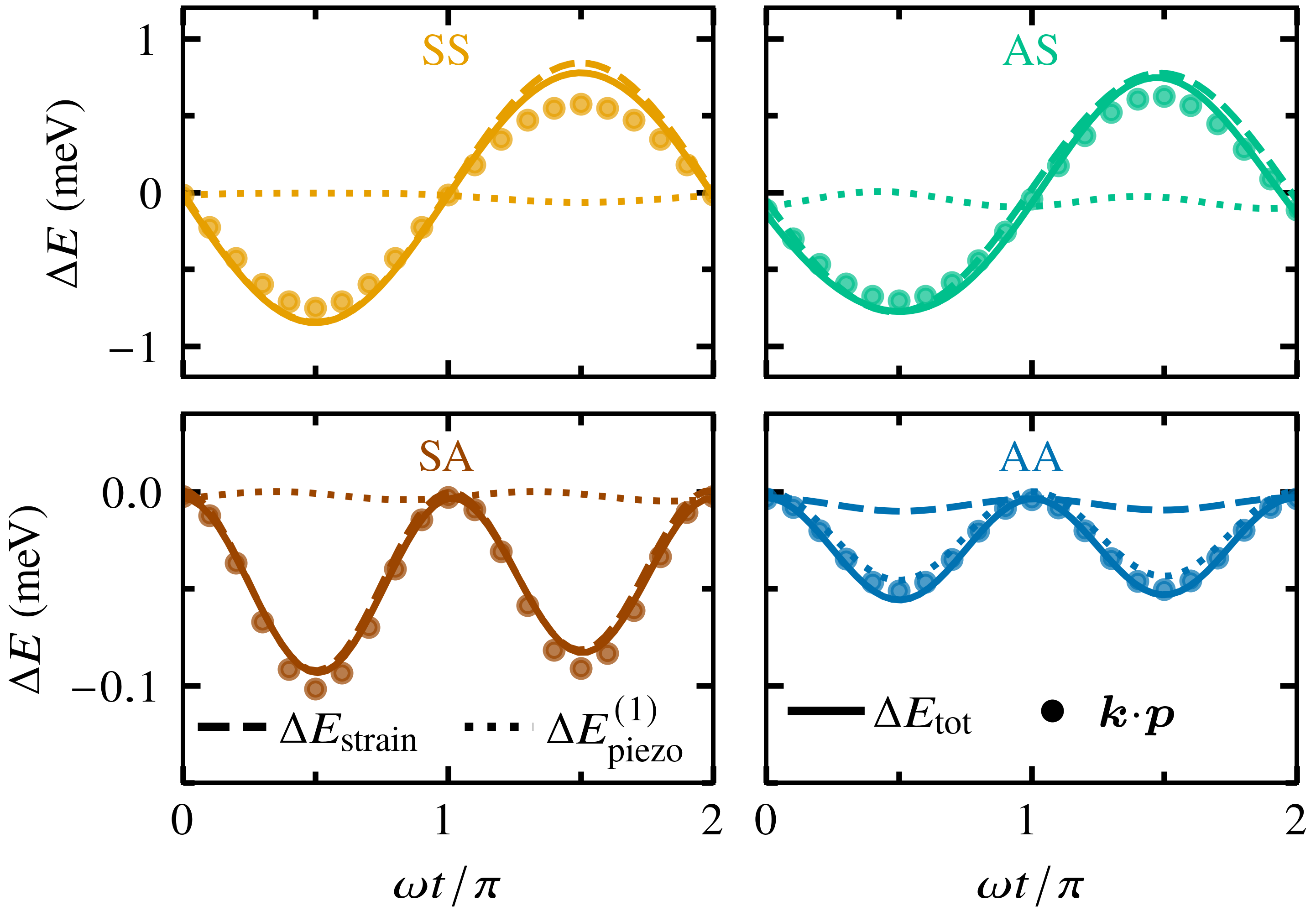}
	\caption{Evolution of all mechanism-induced energy shifts over one oscillation period for all modes: SS (orange), AS (green), SA (brown), and AA (blue). Lines show the results obtained via fitting functions: The contribution from strain, $\Delta E_{\rm{strain}}$, is shown with dashed lines, the piezoelectric dipole contribution, $\Delta E_{\rm{piezo}}^{(1)}$, with dotted lines, and the total shift with solid lines. Points show the results from \kp{}-CI calculations. In both cases, piezoelectric effects are calculated up to the dipole order within the linear polarizability model. Colors distinguish the modes.} \label{fig:fig10-AlGaAs_membrane_TotalEnergyResponse}
\end{figure}  

We calculate the energy contribution from piezoelectric effects using Eq.~\eqref{eq:PiezoResponseDipole} (dotted lines in Fig.~\ref{fig:fig10-AlGaAs_membrane_TotalEnergyResponse}), as before, and the strain-energy contribution (dashed lines in Fig.~\ref{fig:fig10-AlGaAs_membrane_TotalEnergyResponse}) by substituting all six components of the strain tensor into the susceptibility functions from Eqs.~\eqref{eq:diagElements} and~\eqref{eq:fij-GaAs}, allowing the energy shifts to be computed analytically. These two analytically calculated contributions are summed (solid lines in Fig.~\ref{fig:fig10-AlGaAs_membrane_TotalEnergyResponse}) and compared with the \kp{}-CI results, which account for all six strain components and all three electric field components at each time step (points in Fig.~\ref{fig:fig10-AlGaAs_membrane_TotalEnergyResponse}).

For the SS and AS modes, the energy modulation of the QD follows a predominantly harmonic response (upper panel of Fig.~\ref{fig:fig10-AlGaAs_membrane_TotalEnergyResponse}). Both modes are expected to produce a purely sinusoidal energy shift and lead to a modulation amplitude of approximately $0.7$~meV, consistent with previous experimental observations~\cite{vogele2020quantum}.  The energy modulation of the remaining two modes, SA and AA (lower panel of Fig.~\ref{fig:fig10-AlGaAs_membrane_TotalEnergyResponse}) is expected to be about ten times smaller than in the SS and AS cases. Importantly, because the modulation is unidirectional, i.e., the transition energy is only reduced, the resulting time-integrated emission line is predicted to develop in an asymmetric way with the increasing modulation amplitude, with its center shifting towards lower frequencies. This asymmetry arises because the modulation only occurs below the unperturbed transition energy, effectively leading to a low-energy tail in the emission spectrum. So far, experiments using interdigital transducers were configured in such a symmetric way that the excitation predominantly generated AS and SS modes. Consequently, responses associated with AA and SA modes, including frequency doubling, were not observed.

\begin{figure}[tb]
	\includegraphics[width=\columnwidth]{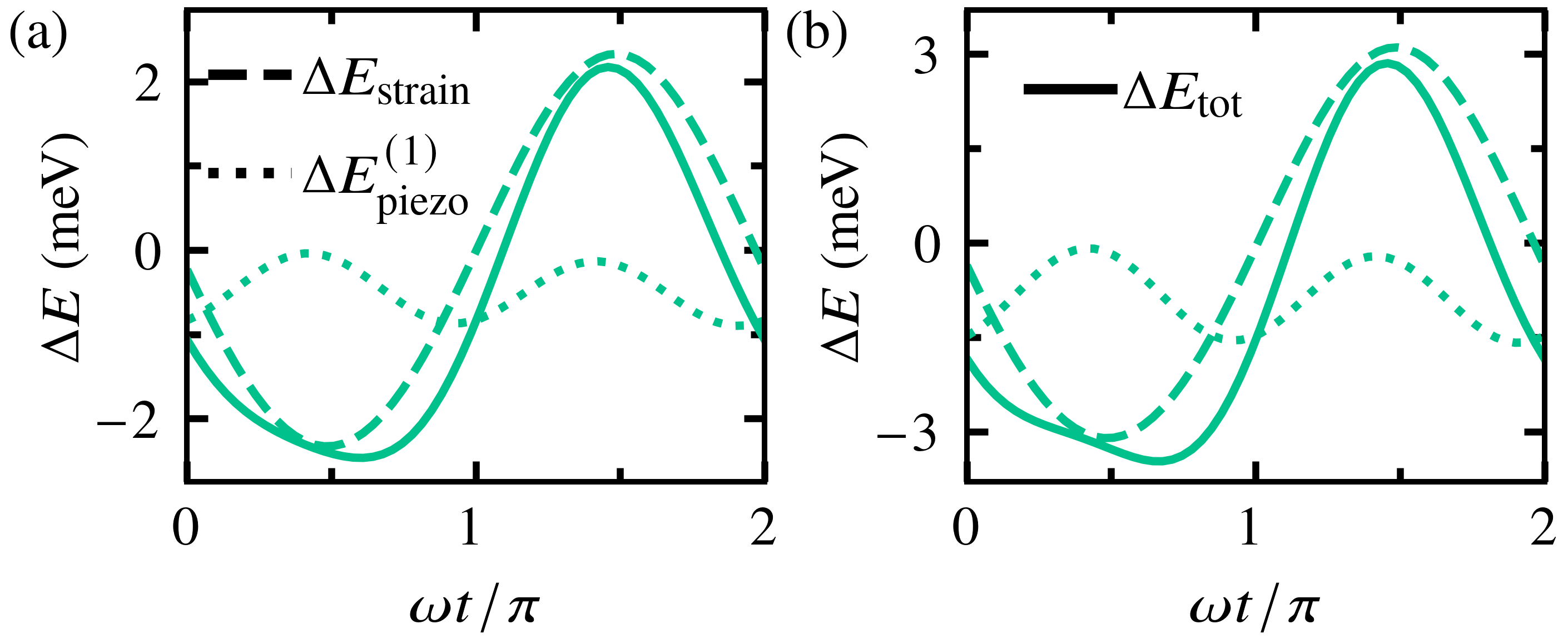}
    \caption{Evolution of all mechanism-induced energy shifts over one oscillation period for the AS mode, for normalization of the total displacement field to 0.3~nm (a) and 0.4~nm (b). The contribution from strain, $\Delta E_{\rm{strain}}$, is shown with dashed lines, and the piezoelectric dipole contribution, $\Delta E_{\rm{piezo}}^{(1)}$, with dotted lines. Solid lines show the total response.}\label{fig:fig11-AlGaAs_membrane_TotalResponse0.3and0.4}
\end{figure}

The AS mode is the only mode that exhibits two coupling channels with different scaling behavior, namely a linear response mediated by volumetric strain and a quadratic response arising from piezoelectric effects. As the fields increase due to a higher amplitude of the mechanical wave, the piezoelectric contribution becomes more significant. This perturbs the nearly perfect sinusoidal response observed for this mode in Fig.~\ref{fig:fig10-AlGaAs_membrane_TotalEnergyResponse} for a normalization amplitude of 0.1~nm. Upon increasing the normalization amplitude to 0.3 and 0.4~nm, this contribution substantially modifies the otherwise sinusoidal energy modulation, as shown in Fig.~\ref{fig:fig11-AlGaAs_membrane_TotalResponse0.3and0.4}. Here, both the piezoelectric and strain contributions were calculated using the linear polarizability model and analytical fitting functions, respectively, since the amplitudes of the electric field and strain tensor components remain within the respective applicability ranges. Such a modified sinusoidal response has been observed experimentally in surface acoustic wave systems~\cite{weiss2014dynamic}.

%%%%%%%%%%%%%%
\section{Conclusions}\label{sec:conclusions}
We have investigated the strain- and electric field-induced modulation of fundamental transition energies of two types of QDs in a PnC membrane engineered to support selected guided acoustic modes within a broad band gap (2.15--3.33\,GHz, with a gap-to-midgap ratio of 43\%). Using \kp{}-CI calculations, we first quantified the intrinsic QD response to individual components of strain and piezoelectric fields. We found that both self-assembled InAs and GaAs QDs grown by droplet etching epitaxy exhibit strong, nearly linear energy shifts under diagonal strain, while shear strain leads to quadratic (second-harmonic) modulation, more pronounced in GaAs QDs due to weaker confinement. Additionally, the dipole and quadrupole moments governing the QD response to electric fields were extracted, revealing a purely built-in vertical dipole in InAs QDs and a tilted one in GaAs QDs. It has been shown that the QD response to in-plane piezoelectric fields is dominated by induced dipoles.

Based on these strain and electric-field susceptibilities, we analyzed the coupling mechanisms of four representative acoustic modes separately according to their symmetry. The SS and AS modes are primarily governed by DP coupling to volumetric strain, with the AS mode exhibiting a pronounced sensitivity to the vertical QD position due to its antisymmetric profile, in contrast to the nearly position-independent SS mode. The SA and AA modes, on the other hand, couple more weakly under the chosen normalization, predominantly via shear strain and piezoelectric effects, respectively.

A detailed analysis of the accompanying piezoelectric fields revealed in-plane electric field amplitudes reaching up to 0.6~mV/nm for a mechanical displacement of 0.1~nm, with a particularly strong enhancement in GaAs membranes owing to their larger piezoelectric coefficients. While the SS mode induces a purely harmonic energy modulation through coupling to the built-in vertical dipole, the AS, SA, and AA modes generate predominantly in-plane fields that act through induced dipole polarizability, giving rise to nonlinear, frequency-doubled responses. Notably, the AS mode gives rise to piezoelectric energy shifts reaching up to 0.18~meV at this displacement amplitude.

When all coupling channels are combined, the SS and AS modes yield the largest overall energy modulation, approaching $\sim$1.5~meV and remaining largely harmonic due to the dominance of DP coupling. In contrast, the SA and AA modes produce an order-of-magnitude smaller modulation, with their asymmetric character arising from dominant piezoelectric and shear-strain interactions, respectively. In the AS mode, the coexistence of comparable deformation-potential and piezoelectric contributions leads to a characteristic non-harmonic modulation of the QD transition energy.

These results provide a comprehensive framework for strain- and electric field-based QD control in phononic nanostructures. Although we implement our approach for waveguides enabling broadband coupling, it can be directly transferred to and applied to the optimization of PnC cavities~\cite{hatanaka2020real}. High phononic quality factors and small mode volumes pave the way for efficient, Purcell-enhanced single-phonon generation and potentially quantum transduction in the strong-coupling regime. Moreover, direct transfer from the considered snowflake-type PnC platform to other geometries is straightforward. It is noteworthy that the considered snowflake platform also supports tightly confined photonic modes~\cite{safavi2014two}. These, strongly coupled to the phononic component considered here, then allow full-fledged optomechanical transduction by the QD. This universality ultimately opens new strategies and avenues for the design of hybrid quantum devices and phonon-mediated quantum transduction.

\begin{acknowledgements}
We are grateful to Armando Rastelli for discussions on QD growth and images of nanoholes forming GaAs QDs. The authors thank Benjamin Mayer for valuable discussions during the initial stages of this work.
M.~G. is grateful to Krzysztof Gawarecki for sharing his computational code.
This project was supported by the German Federal Ministry of Education and Research via the Research Group Linkage Program of the Alexander von Humboldt Foundation. J.~R. and P.~M. acknowledge funding from the Narodowe Centrum Nauki (NCN, Polish National Science Centre), grant no. 2023/50/A/ST3/00511. M.~W. and H.~K. acknowledge support by the Deutsche Forschungsgemeinschaft (DFG, German Research Foundation), grant no. 465136867. 
M.~G. acknowledges the financing of the MEEDGARD project funded within the QuantERA II Program that has received funding from the European Union's Horizon 2020 research and innovation program under Grant Agreement No.\ 101017733 and the National Centre for Research and Development, Poland -- project No.\ QUANTERAII/2/56/MEEDGARD/2024.
Part of the numerical calculations has been carried out using resources provided by Wroclaw Centre for Networking and Supercomputing (http://wcss.pl), Grant No. 203.

\end{acknowledgements}

\appendix

\section{Material parameters and QD geometry}
\label{sec:mat-params}
			\begingroup
			\begin{table}[]
				\newcommand{\cw}{\cite{Winkler2003}}
				\newcommand{\cs}{\cite{SaidiJAP2010}}
                \newcommand{\ccaro}{\cite{Caro2015}}
				\newcommand{\ca}{\cite{AmirtharajBOOK1994}}
                \newcommand{\cvlad}{\cite{Mlinar2005}}
                \newcommand{\clb}{\cite{Landolt-Bornstein}}
				\newcommand{\cmsq}{${\mathrm{C}}/{\mathrm{m}^2}$}
				\newcommand{\intEg}{$\substack{-0.13\\+1.31x}$}
				\begin{tabular*}{0.48\textwidth}{@{\extracolsep{0.1cm}}l|lllll}
					\toprule\rule{0pt}{1.1em} 
	&	AlAs	&	GaAs	&	InAs	&	$C^{\mathrm{AlAs}}_{\mathrm{GaAs}}$	&	$C^{\mathrm{GaAs}}_{\mathrm{InAs}}$	\\[2pt]
     \hline\rule{0pt}{1.1em}											
     $a_0$\,(\AA)    	&	5.652 	&	5.642 	&	6.05 	&	0  	&	0  	\\
     $E_{\mathrm{g}}$\,(eV)   	&	3.099 	&	1.519 	&	0.417 	&	\intEg 	&	0.477 	\\
     VBO\,(eV)      	&	$-1.32$ 	&	$-0.80$ 	&	$-0.59$ 	&	0  	&	$-0.38$ 	\\
     $m_{\mathrm{e}}^{*}$     	&	0.15 	&	0.0665 	&	0.0229 	&	0  	&	0.0091 	\\
     $\Delta_{\rm SO}$\,(eV)   	&	0.28 	&	0.341 	&	0.39 	&	0  	&	0.15 	\\
     $\gamma_{\mathrm{1}}$  	&	3.76 	&	6.98 	&	20.0 	&	0  	&	0  	\\
     $\gamma_{\mathrm{2}}$  	&	0.82 	&	2.06 	&	8.5  	&	0  	&	0  	\\
     $\gamma_{\mathrm{3}}$	&	1.42 	&	2.93 	&	9.2  	&	0  	&	0  	\\
     $e_{\mathrm{14}}$\,(\cmsq)\ccaro   	&	$-0.055$ 	&	$-0.205$ 	&	$-0.111$ 	&	0  	&	0  	\\
     $B_{\mathrm{114}}$\,(\cmsq)\ccaro  	&	$-1.61$  	&	$-0.99$ 	&	$-1.17$ 	&	0  	&	0  	\\
     $B_{\mathrm{124}}$\,(\cmsq)\ccaro  	&	$-2.59$  	&	$-3.21$ 	&	$-4.31$ 	&	0  	&	0  	\\
     $B_{\mathrm{156}}$\,(\cmsq)\ccaro  	&	$-1.32$  	&	$-1.28$ 	&	-0.46  	&	0  	&	0  	\\
     $C_{\mathrm{k}}$\,(eV\AA)    	&	0.002 	&	$-0.0034$ 	&	$-0.0112$ 	&	0  	&	0  	\\
     $a_{\mathrm{c}}$\,(eV)    	&	$-5.64$ 	&	$-7.17$ 	&	$-5.08$ 	&	0  	&	2.61 	\\
     $a_{\mathrm{v}}$\,(eV)    	&	2.47 	&	1.16 	&	1.0  	&	0  	&	0  	\\
     $b_{\mathrm{v}}$\,(eV)    	&	$-2.3$ 	&	$-2.0$ 	&	$-1.8$ 	&	0  	&	0  	\\
     $d_{\mathrm{v}}$\,(eV)    	&	$-3.4$ 	&	$-4.8$ 	&	$-3.6$ 	&	0  	&	0  	\\
     $C_{11}$\,(GPa)   	&	125.0 	&	122.1 	&	83.3  	&	0  	&	0  	\\
     $C_{12}$\,(GPa)    	&	53.4 	&	56.6 	&	45.3  	&	0  	&	0  	\\
     $C_{44}$\,(GPa)    	&	54.2 	&	60.0 	&	39.6  	&	0  	&	0  	\\
     $\kappa$\cw   	&	10.06 	&	12.4 	&	14.6 	&	0  	&	0  	\\
     $\rho$\,(kg/m$^{3})$\cite{adachi1985gaas}   &	3760  & 5360 &	 	&	0  	&	0  	\\
					\toprule
				\end{tabular*}
				\caption{\label{tab:params} Material parameters used in the modeling of QDs and calculation of single-particle and exciton states; $C^\mathrm{A}_{\mathrm{B}}$ are values of ternary bowing parameters. Unless otherwise marked, parameters are taken from Ref.\,\onlinecite{vurgaftman2001band}.}
			\end{table}
			\endgroup

Table~\ref{tab:params} lists all the material parameters for InAs, GaAs, and AlAs used in our simulations. For InGaAs and AlGaAs alloys, we use linear interpolation, with additional bowing terms included as indicated in the table. The parameters come from Ref.~\onlinecite{vurgaftman2001band} unless otherwise noted. For the Kane energy, we used the formula
\begin{equation*}
    E_{P} = \left(\frac{m_{0}}{m_{\mathrm{e}}^{*}} - 1 \right) \frac{E_{\mathrm{g}}(E_{\mathrm{g}}+\Delta_{\rm SO})}{(E_{\mathrm{g}}+2\Delta_{\rm SO}/3)}
\end{equation*}
to preserve ellipticity of the \kp{} equation system for envelope functions~\cite{VeprekPRB2007,BirnerBOOK2014}.

\begin{figure}[ht]
    \centering
    \includegraphics[width=\linewidth]{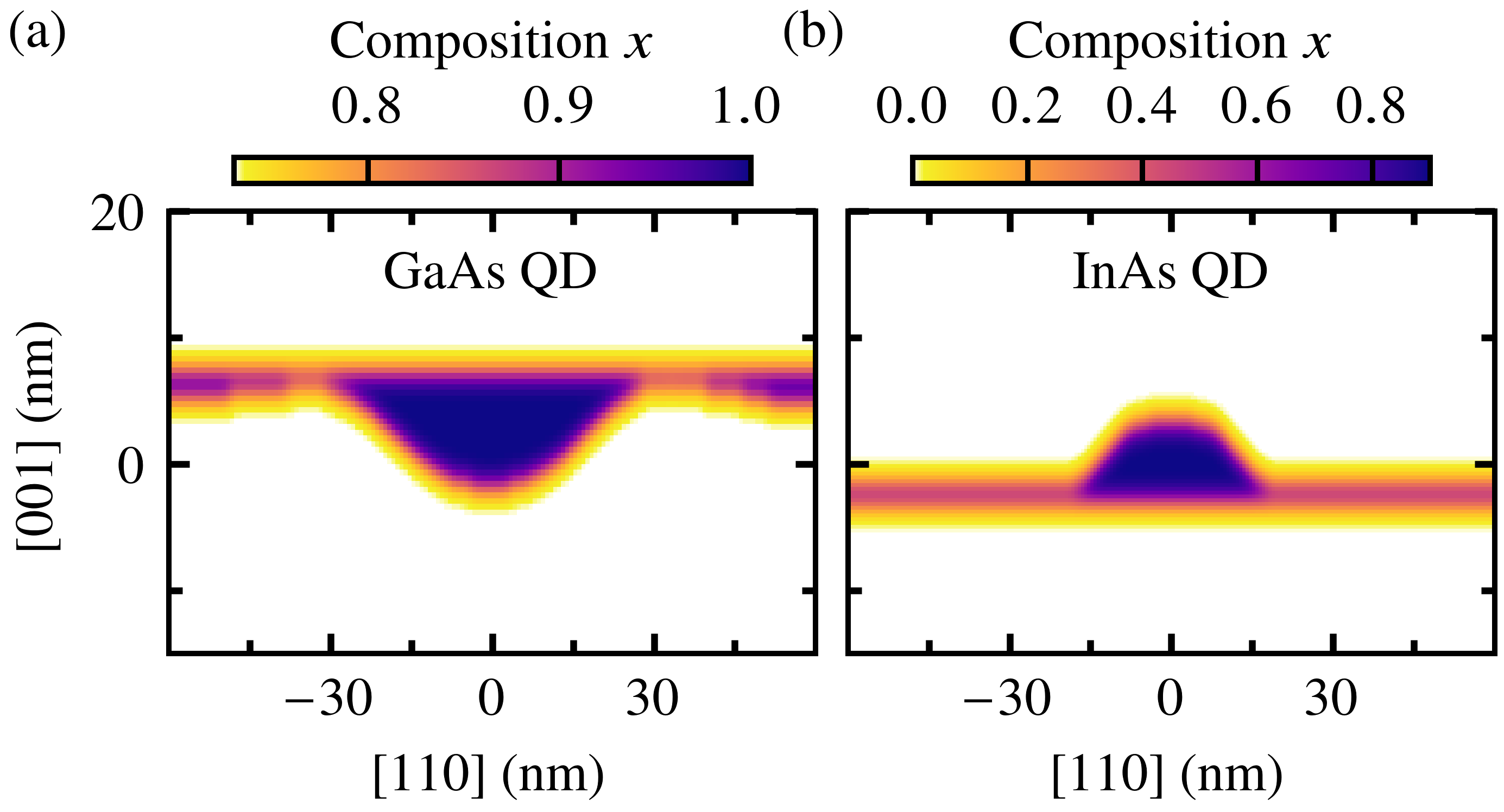}
    \caption{The model of the investigated QDs. The color scale shows: (a) the gallium content $x$ in the Al$_{1-x}$Ga$_x$As material, (b) the indium content $x$ in the In$_x$Ga$_{1-x}$As material.}
    \label{fig:fig12-map_QDs}
\end{figure}

Figure~\ref{fig:fig12-map_QDs}(a) shows the cross section of the modeled dome-shaped InAs/GaAs QD with In$_{0.9}$Ga$_{0.1}$As QD material based on prior work~\cite{Ardelt2016}, with color-coded material composition, and Fig.~\ref{fig:fig12-map_QDs}(b) presents a similar view of the GaAs/AlGaAs QD based on the atomic force microscopy image of a representative nanohole forming a QD~\cite{Yuan2023,schimpf2025optical}.

The InAs QD has an elliptical base with major and minor axes of 44~nm and 38~nm along the [110] and [1$\bar{1}0$] directions, respectively, and a height of 5~nm. The GaAs QD has a nearly conical shape with a base diameter of approximately 60~nm and a height of 9~nm.

\section{Acoustic dispersion curves for AlGaAs waveguide}\label{sec:AlGaAs}
\begin{figure}[ht]
    \centering
    \includegraphics[width=\linewidth]{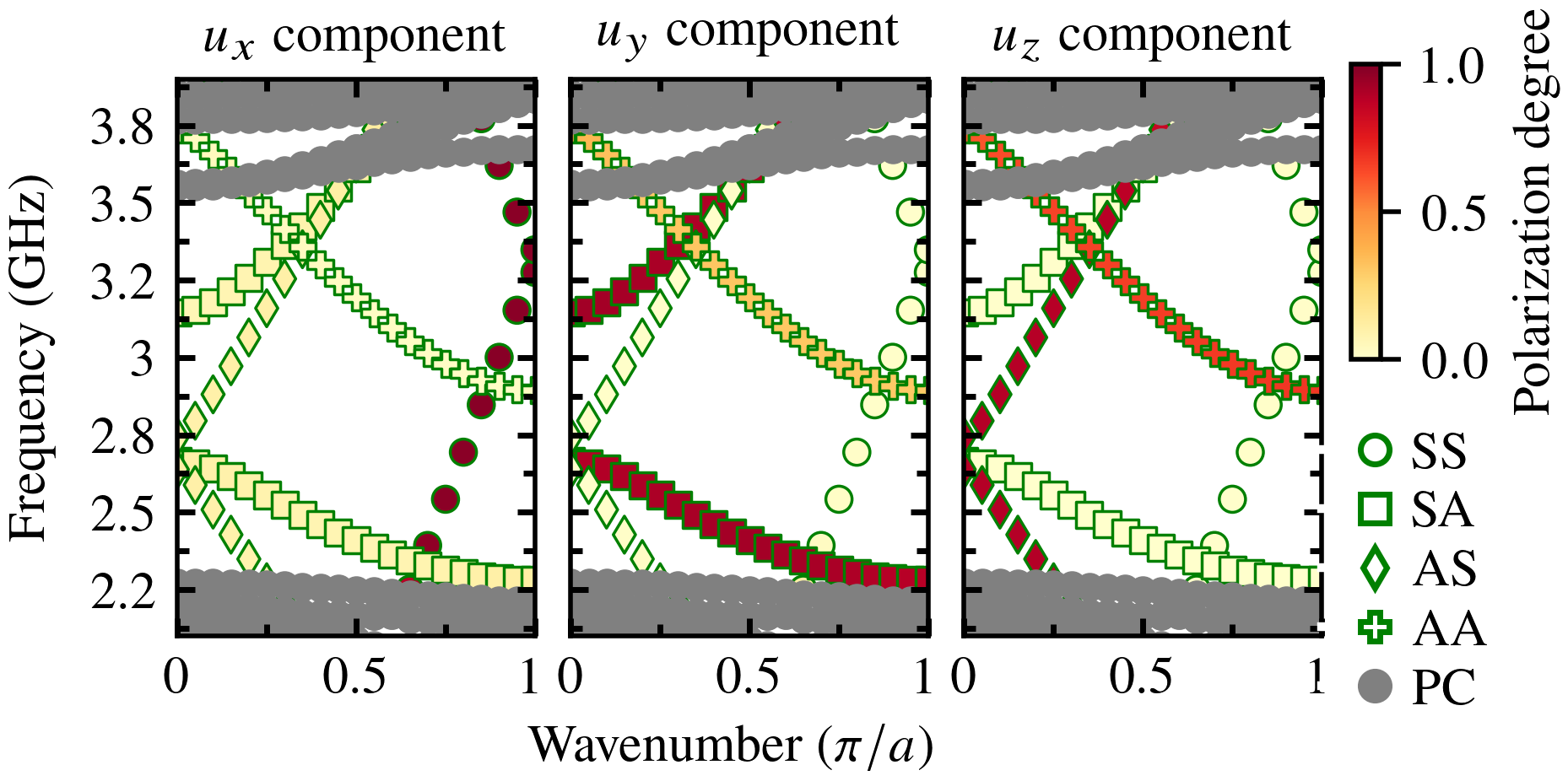}
    \caption{Dispersion curves of guided modes in a W1m waveguide based on Al$_{0.4}$Ga$_{0.6}$As with $\Delta = 0.6a$. Point shapes indicate the symmetry of the mode. The color scale shows the contribution of the respective displacement components, $P_l$, averaged over the supercell volume, according to Eq.~\eqref{eq:Pl} in the main text.}
    \label{fig:fig13-w1m_delta060_AlGaAs_skladowe}
\end{figure}
Fig.~\ref{fig:fig13-w1m_delta060_AlGaAs_skladowe} shows the acoustic band structure of a phononic waveguide made in an AlGaAs membrane. These bands are calculated for the same geometry (identical parameters $a$, $r$, $w$, $d$, $\Delta$) as the GaAs membrane shown in Fig.~\ref{fig:fig4-skladoweCombined}(b) in the main text. The dispersion curves and polarization distributions are identical to those in the GaAs waveguide. The mode profiles remain identical, differing only by a shift in their operational frequencies. Due to its lower mass density, the AlGaAs waveguide exhibits the same dispersion relations shifted upward in frequency by approximately 200 MHz.

\section{Single-particle response to strain}\label{sec:SPB}
\begin{figure}[tb]
	\includegraphics[width=\columnwidth]{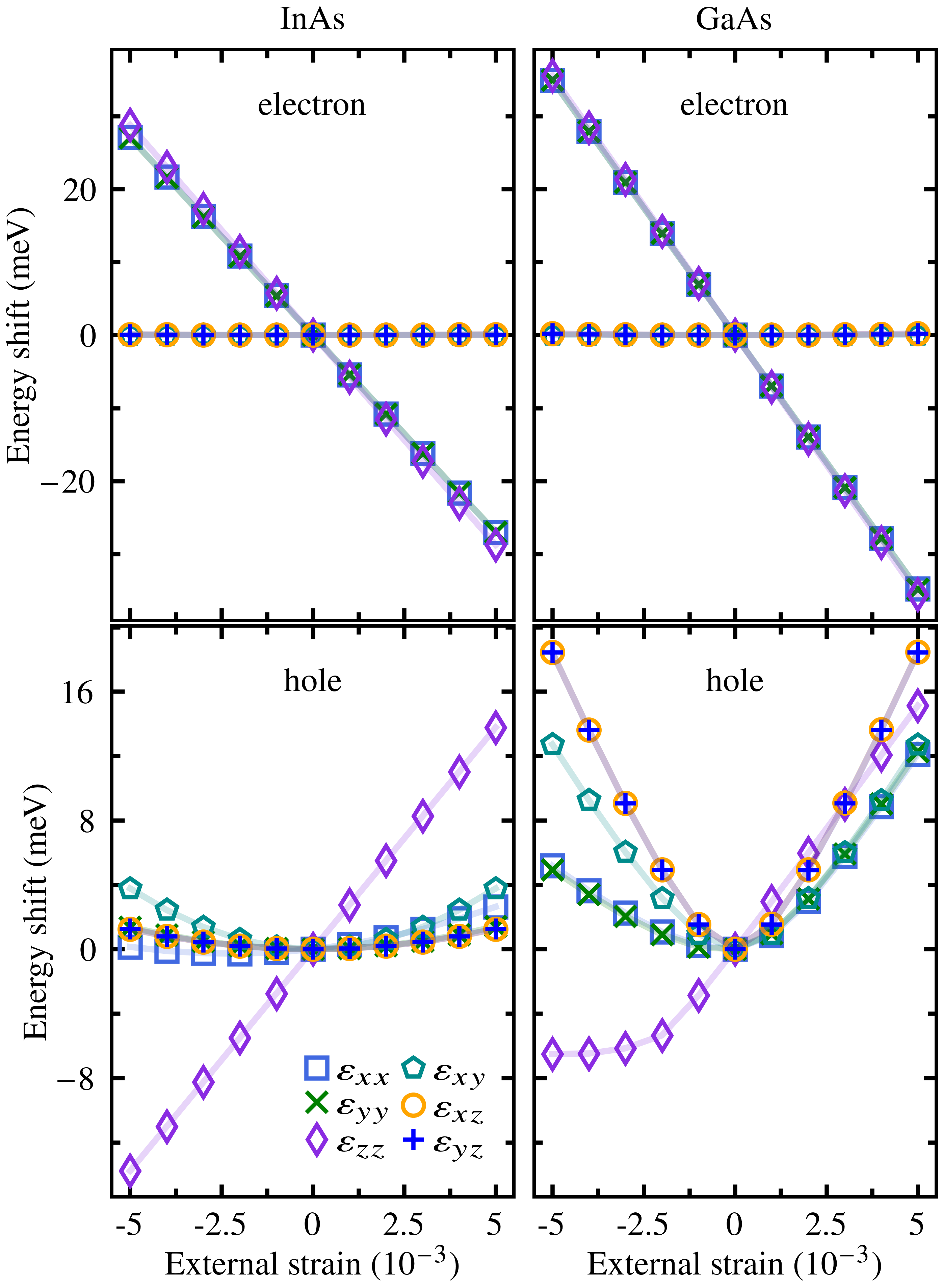}
	\caption{Energy modulation of the ground-state electron (upper row) and hole (lower row) in the InAs (left column) and GaAs (right column) QDs as a function of different external strain components. The color of each curve corresponds to the different strain component.}
    \label{fig:fig14-eh-both-InAs-GaAs}
\end{figure}

Figure~\ref{fig:fig14-eh-both-InAs-GaAs} shows the DP effect calculated directly using our \kp{} model for the ground-state electron and hole single-particle states in a QD subjected to a single-component external strain. In both QDs, electrons respond exclusively to the diagonal components of the strain tensor in a linear manner. The hole response is nonlinear for all strain components except the $\varepsilon_{zz}$ component in the InAs QD. This results in an almost perfectly linear excitonic response in the InAs QD, whereas the GaAs QD exhibits noticeable nonlinearity even for $\varepsilon_{zz}$ strain. Shear strain induces a clearly parabolic energy shift that is significantly stronger in the GaAs QD.

\section{Quadrupole corrections}
\label{appendix:quadrupole}

In this Appendix, we discuss the convergence of the multipole expansion and show that the next order, i.e., the quadrupole corrections are negligible in our system. In general, since the electric field varies spatially on a scale of the supercell size $a$ or the membrane thickness $d$ and the charge distribution is confined to the QD size $L$, the convergence of the multipole expansion is governed by the parameters $L/a$ or $L/d$, both of which are at most on the order of 0.1. We therefore expect the quadrupole contribution to be negligible, as confirmed below.

\begin{table}[tb]
\begin{ruledtabular}
\begin{tabular}{lcc}
\textrm{Component (nm$^2e$)} & \textrm{InAs} & \textrm{GaAs} \\
\colrule
$Q_{\tilde{x}\tilde{x}}$  & $14.88$  & $-0.01953$  \\
$Q_{\tilde{y}\tilde{y}}$  & $-4.103$  & $-3.676$  \\
$Q_{\tilde{z}\tilde{z}}$ & $-10.78$  & $3.696$  \\
\end{tabular}
\end{ruledtabular}
\caption{\label{tab:quadrupol}%
Principal values of the quadrupole moments for InAs and GaAs QDs at null electric field. $\tilde{x},\tilde{y},\tilde{z}$ denote the principal axes, which turn out to be very close to the axes of our reference frame.}
\end{table}

We calculate the quadrupole moments according to Eq.~\eqref{eq:dip_Quad} using the charge distributions found from the \kp{}-CI calculations in the same way as for the dipole moments, as described in Sec.~\ref{subsec:kp_calc}. Next, the quadrupole moment tensor is diagonalized to find its principal axes and principal values. The principal values of the built-in quadrupole moment (i.e., at null electric field) for both QD systems studied in this work are listed in Tab.~\ref{tab:quadrupol}. The principal axes of the built-in quadrupole nearly coincide with the crystallographic axes along which the waveguide is aligned: the tilt between the principal axes and the corresponding axes $x,y,z$ of our model is below $3^{\circ}$ and below $6^{\circ}$ for the InAs and GaAs QD, respectively. We denote these principal directions by $(\tilde{x},\tilde{y},\tilde{z})$. Data files supporting this analysis are available as Supplementary Material to this paper~\cite{sm2026}. As can be seen from the field profiles presented in Appendix~\ref{appendix:ModeProfiles}, the strongest field gradients do not exceed 7.5~$\mu$V/nm$^2$. For the built-in quadrupole moments found here, not exceeding 15~$e\cdot$nm$^2$, this leads to the upper bound for the energy shifts on the order of 0.01~meV, well below the other effects.

\begin{figure}[tb]
	\includegraphics[width=\columnwidth]{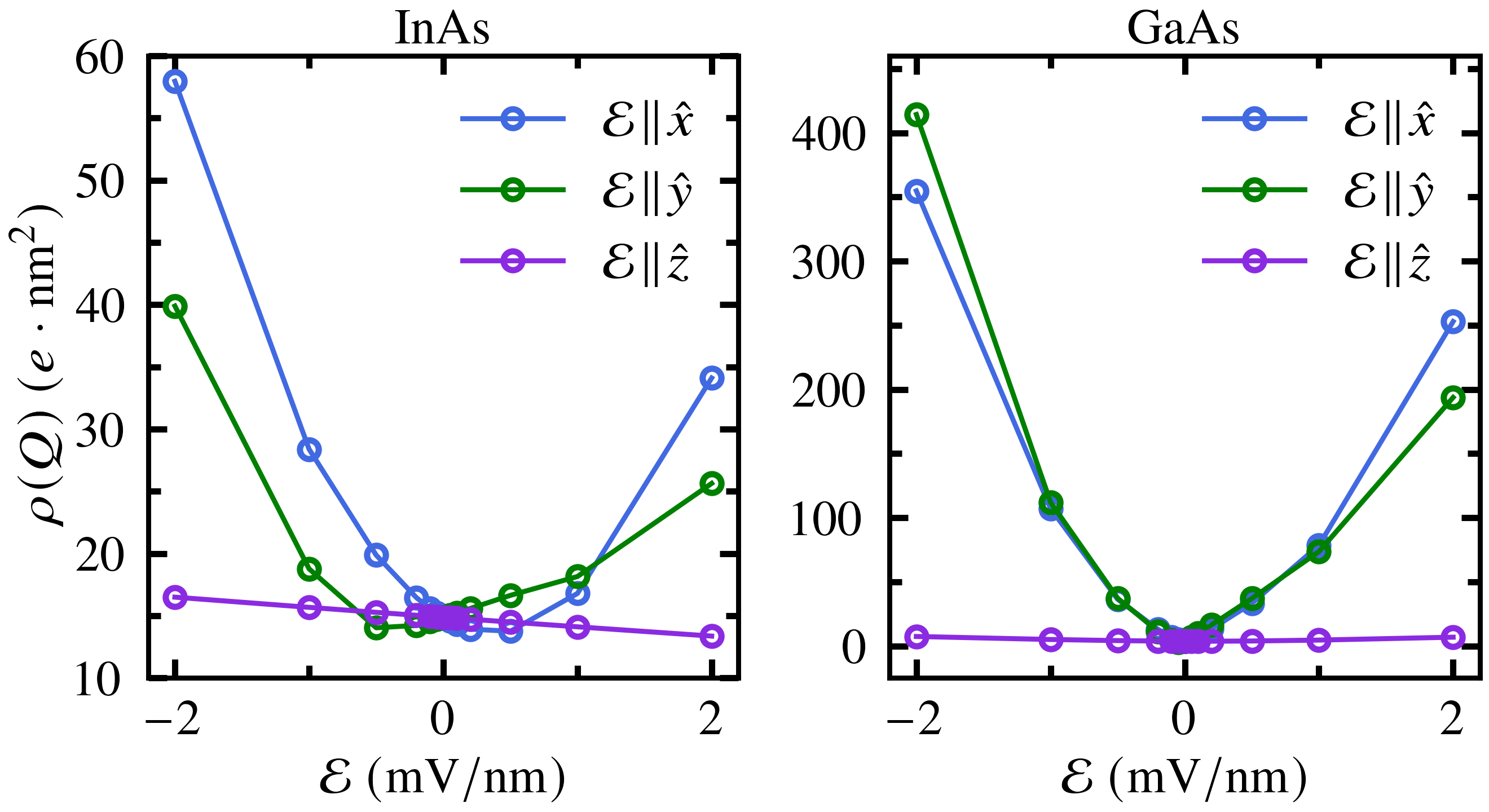}
	\caption{The dependence of the magnitude (highest principal value) of the electric quadrupole moment for the InAs (left) and GaAs (right) QD on the electric field applied along the axes of the reference frame. Lines are added as a guide to the eye.} 
    \label{fig:fig15-Quad_Response}
\end{figure}

As we have seen in the case of dipole moments, electric fields of magnitudes typical for our system considerably modify the charge distribution. This also affects the quadrupole moments of the QD charge distributions, as we show in Fig.~\ref{fig:fig15-Quad_Response} where, for the sake of estimating the upper bound on the effect, we only plot the highest modulus of the quadrupole principal value (that is, the spectral range of the quadrupole tensor, $\rho(Q)$) as a function of the electric field applied along the three axes of our coordinate system. The induced quadrupole moments turn out to be quite large, in particular for the GaAs QD, but only in response to an in-plane electric field. At the highest fields achievable here (about 0.6~mV/nm), the induced quadrupole moment might reach 40~$e\cdot$nm$^2$, leading to energy modulation up to 0.05~meV. This is still a small value and represents a crude upper bound, which is actually overestimated for each specific case. The strongest fields and the strongest gradients appear for different modes. Moreover, even when strong in-plane fields coexist with large gradients, like for the AS mode [see Fig.~\ref{fig:fig17-all_modes_GaAs_Piezo}(b)], these two quantities reach their maxima at different spatial points. 

For significantly larger acoustic amplitudes, the quadratic (or even cubic) scaling of the induced-quadrupole contribution could increase its relative importance. At the same time, however, the induced-dipole term scales identically and is expected to remain dominant.

\section{Strain and electric field profiles}\label{appendix:ModeProfiles}

In this Appendix, we show and further discuss the strain tensor profiles for the SS, SA, AS, and AA modes of a GaAs membrane waveguide studied in the text.

\begin{figure*}[p]
    \centering
    \includegraphics[width=1.0\linewidth]{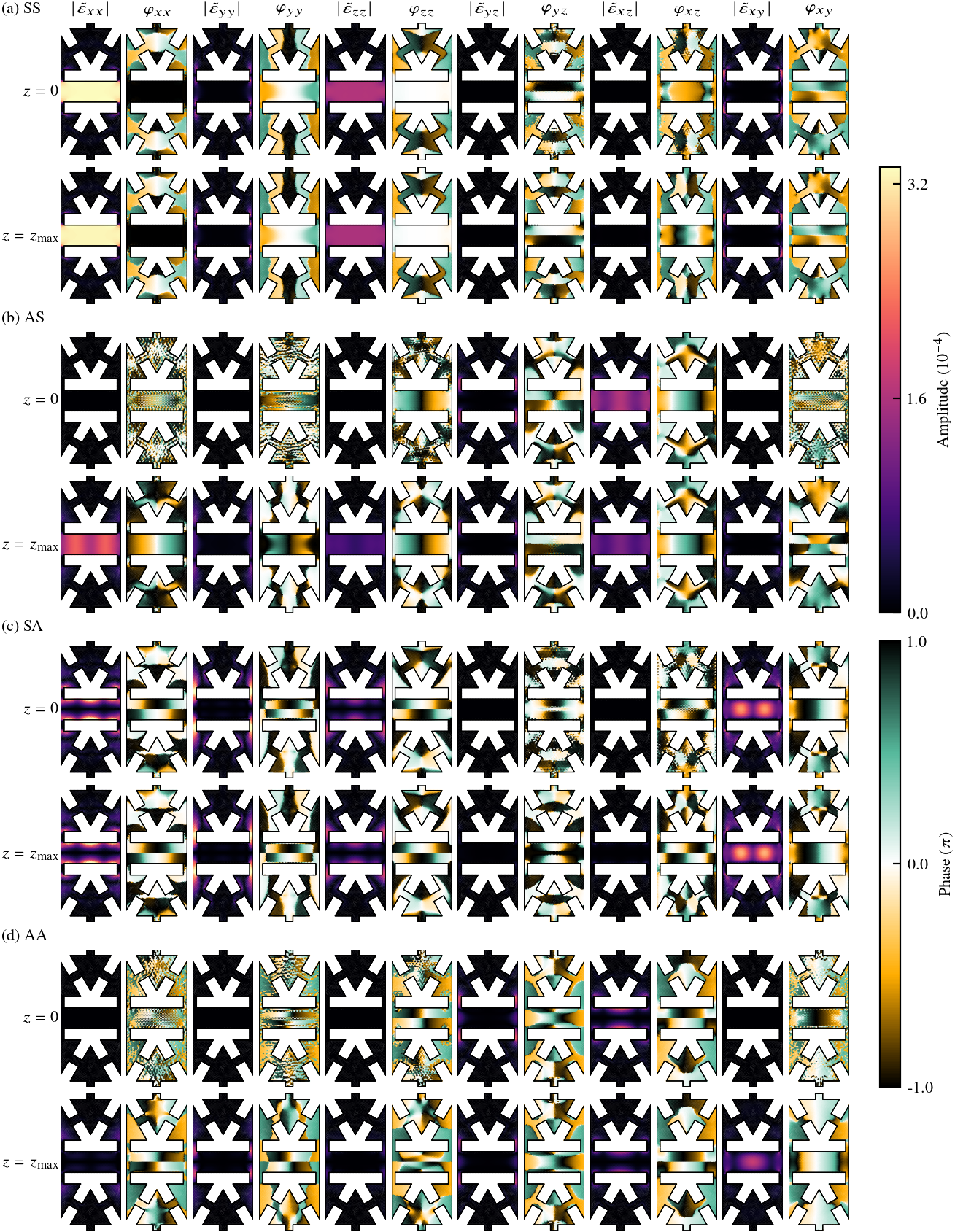}
    \caption{Strain tensor components for (a) SS, (b) AS, (c) SA, and (d) AA modes in a GaAs membrane, corresponding to $(ak/\pi,f) = (0.8,2.54\,\text{GHz})$, $(0.15,2.30\,\text{GHz})$, $(0.15,3.00\,\text{GHz})$, and $(0.8,2.78\,\text{GHz})$, respectively. Results are shown in two parallel planes: $z = 0$ (upper panel) and $z = z_{\rm max}$ (lower panel). Each panel displays all six independent components of the strain tensor, with each component represented by an amplitude (left) and phase (right).}
    \label{fig:fig16-all_modes_GaAs_StrainsV}
\end{figure*}

Fig.~\ref{fig:fig16-all_modes_GaAs_StrainsV} shows the amplitudes and phases of the six independent strain tensor components for the four modes discussed in the main text. The profiles are plotted in the $xy$ plane for two cross-sections along the thickness direction (at $z=0$ and $z=z_{\rm max})$ to capture both in-plane and out-of-plane variations. This allows a complete characterization of the strain fields and their symmetry properties.

In the SS mode [Fig.~\ref{fig:fig16-all_modes_GaAs_StrainsV}(a)], the volumetric strain arises from the \(\varepsilon_{xx}\) and \(\varepsilon_{zz}\) components. Their opposite signs reduce the overall volumetric strain amplitude. The strain distribution is approximately independent of $z$ due to the symmetry of the mode. No shear strain components are present in that mode.

The volumetric strain of the AS mode [Fig.~\ref{fig:fig16-all_modes_GaAs_StrainsV}(b)] is driven by the same components as the SS mode, \(\varepsilon_{xx}\) and \(\varepsilon_{zz}\), also with opposite signs. Both components exhibit vertical dependence, vanishing at \(z = 0\). Shear strain coupling is enabled by the \(\varepsilon_{xz}\) component, which is strongest at \(z = 0\) and decreases to zero toward the external plane at \(z = d/2\). The vertical dependence of both volumetric and shear strains arises from the mode's asymmetry with respect to the \(\sigma_y\) plane.

The SA mode [Fig.~\ref{fig:fig16-all_modes_GaAs_StrainsV}(c)] does not exhibit volumetric strain in the region of interest (at least 50~nm from the interfaces) due to the mode’s asymmetry with respect to the \( \sigma_y \) plane. This plane enforces the cancellation of all strain components except for the \( \varepsilon_{xy} \) component at \( y = 0 \) for any \( x \) and \( z \), which pushes the strain maxima outward from the core toward the PnC. The shear strain component \( \varepsilon_{xy} \) remains vertically uniform due to symmetry.

The AA mode [Fig.~\ref{fig:fig16-all_modes_GaAs_StrainsV}(d)], antisymmetric with respect to both planes, is similar to the SA mode—no volumetric strain is present due to the asymmetry with respect to the \( \sigma_y \) plane. 
%Coupling occurs via the \( \varepsilon_{xy} \) shear strain component, enabling interaction with both QD types, though again it is weak for the InAs QD. 
The shear strain exhibits vertical dependence (similar to the AS mode), which distinguishes it from the SA mode and results from the mode's asymmetry with respect to the \( \sigma_z \) plane.

The strain distribution in the waveguide modes is strongly influenced by their symmetry properties. Volumetric strain is significant in modes symmetric with respect to the \(\sigma_y\) plane (SS and AS) but absent in modes antisymmetric with respect to that plane (SA and AA). The first pair of modes can be efficiently converted from bulk Rayleigh waves into the waveguide structures studied here~\cite{korovin2019conversion}. Shear strain, on the other hand, plays a crucial role in antisymmetric modes, with \(\varepsilon_{xy}\) and \(\varepsilon_{xz}\) components enabling coupling with QDs. The vertical dependence of strain components is directly linked to the mode's symmetry with respect to the \(\sigma_z\) plane: symmetric modes (SS, SA) exhibit no vertical dependence, while antisymmetric modes (AS, AA) show clear vertical variations. These findings highlight the importance of symmetry in tailoring strain-induced interactions for optomechanical applications.

\begin{figure*}[p]
    \centering
    \includegraphics[width=\linewidth]{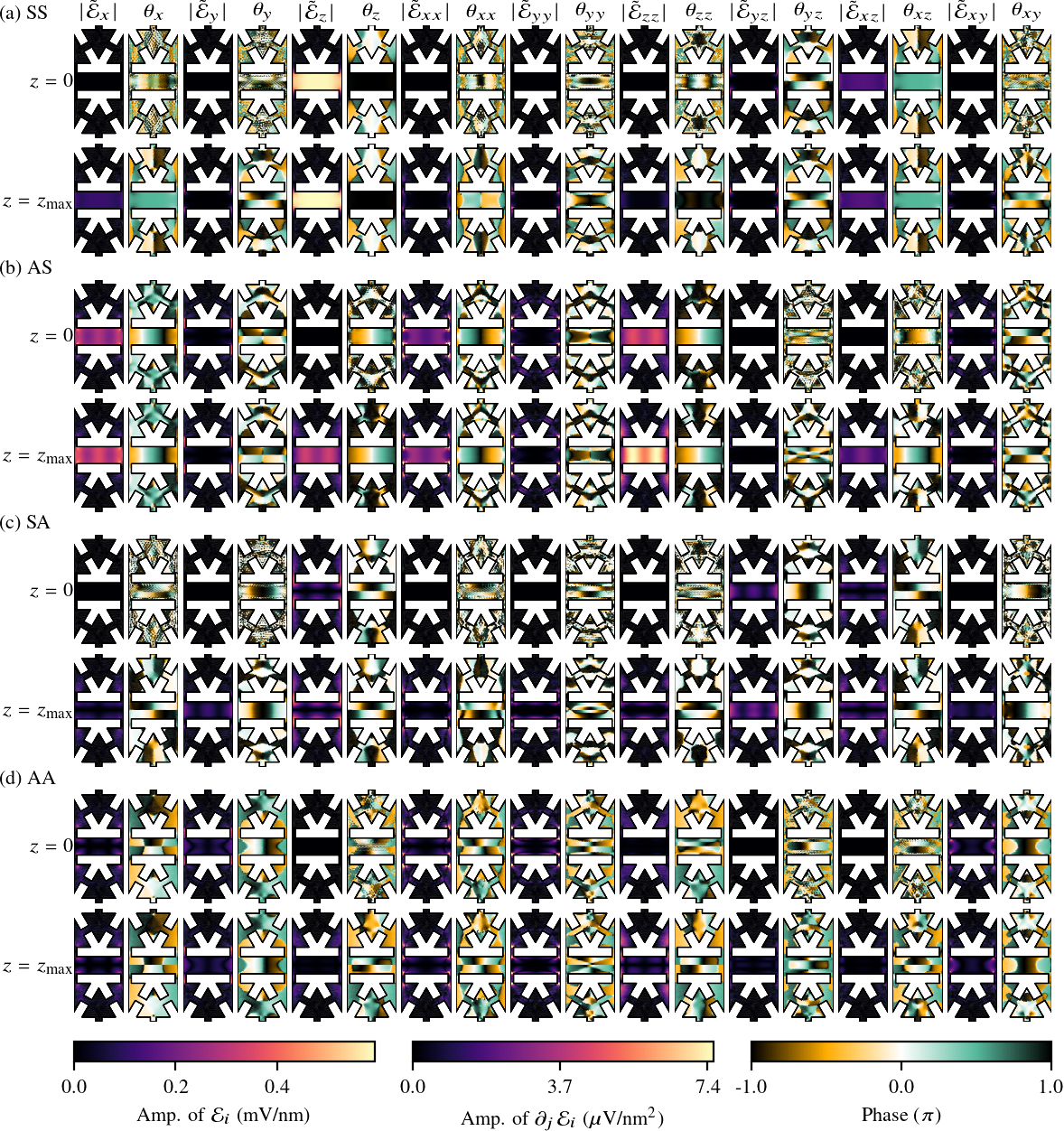}
    \caption{Electric field components and their spatial derivatives for (a) SS, (b) AS, (c) SA, and (d) AA modes in a GaAs membrane, corresponding to $(ak/\pi,f) = (0.8,2.54\,\text{GHz})$, $(0.15,2.30\,\text{GHz})$, $(0.15,3.00\,\text{GHz})$, and $(0.8,2.78\,\text{GHz})$, respectively. Results are shown in two parallel planes: $z = 0$ (upper panel) and $z = z_{\rm max}$ (lower panel). Each panel displays nine field quantities, with each quantity represented by a pair of plots showing amplitude (left) and phase (right). The first three pairs correspond to electric field components, followed by six pairs showing field derivatives.}
    \label{fig:fig17-all_modes_GaAs_Piezo}
\end{figure*}

Fig.~\ref{fig:fig17-all_modes_GaAs_Piezo} presents the magnitudes and phases of the components of electric fields and their gradients, resulting from the piezoelectric effect. The fields are shown on the same two horizontal planes as in the previous figures.  

Modes symmetric under reflection with respect to the $z = 0$ plane ($\sigma_z$), SS and SA, exhibit diagonal and in-plane shear ($\varepsilon_{xy}$) strains that are symmetric, while the out-of-plane shear components ($\varepsilon_{xz}$, $\varepsilon_{yz}$) and the electric potential are antisymmetric. This results in an approximately uniform $\mathcal{E}_z$ component along $z$ and vanishing in-plane components $\mathcal{E}_x$, $\mathcal{E}_y$ in the symmetry plane ($z = 0$) for these modes [see Fig.~\ref{fig:fig17-all_modes_GaAs_Piezo}(a) and Fig.~\ref{fig:fig17-all_modes_GaAs_Piezo}(c)].

In contrast, modes that are anti-symmetric under $\sigma_z$ (AS, AA) behave oppositely: the potential is symmetric, leading to a more uniform in-plane field distribution and an antisymmetric $\mathcal{E}_z$ component that vanishes in the $z = 0$ plane.

The SS and SA modes [Figs.~\ref{fig:fig17-all_modes_GaAs_Piezo}(a) and (c)] exhibit a dominant out-of-plane piezoelectric field. The SA mode also features a $\mathcal{E}_y$ component that increases with vertical position. The AS mode [Fig.~\ref{fig:fig17-all_modes_GaAs_Piezo}(b)] exhibits both a strong in-plane piezoelectric field ($\mathcal{E}_x$) and an out-of-plane component ($\mathcal{E}_z$). The AA mode [Fig.~\ref{fig:fig17-all_modes_GaAs_Piezo}(d)] features both in-plane piezoelectric field components ($\mathcal{E}_x$ and $\mathcal{E}_y$).

In AlGaAs membranes, the strain fields maintain identical modal profiles and magnitudes compared to GaAs. The piezoelectric fields, however, show markedly different behavior due to AlGaAs's significantly lower piezoelectric tensor $e_{ijk}$, resulting in approximately 30\% weaker fields in the AlGaAs membrane.

\FloatBarrier
\bibliography{paper}

% ---------- wstawianie dipoli (3 strony) ----------
\foreach \i in {1,2,3} {%
  \clearpage
  \includepdf[pages={\i}]{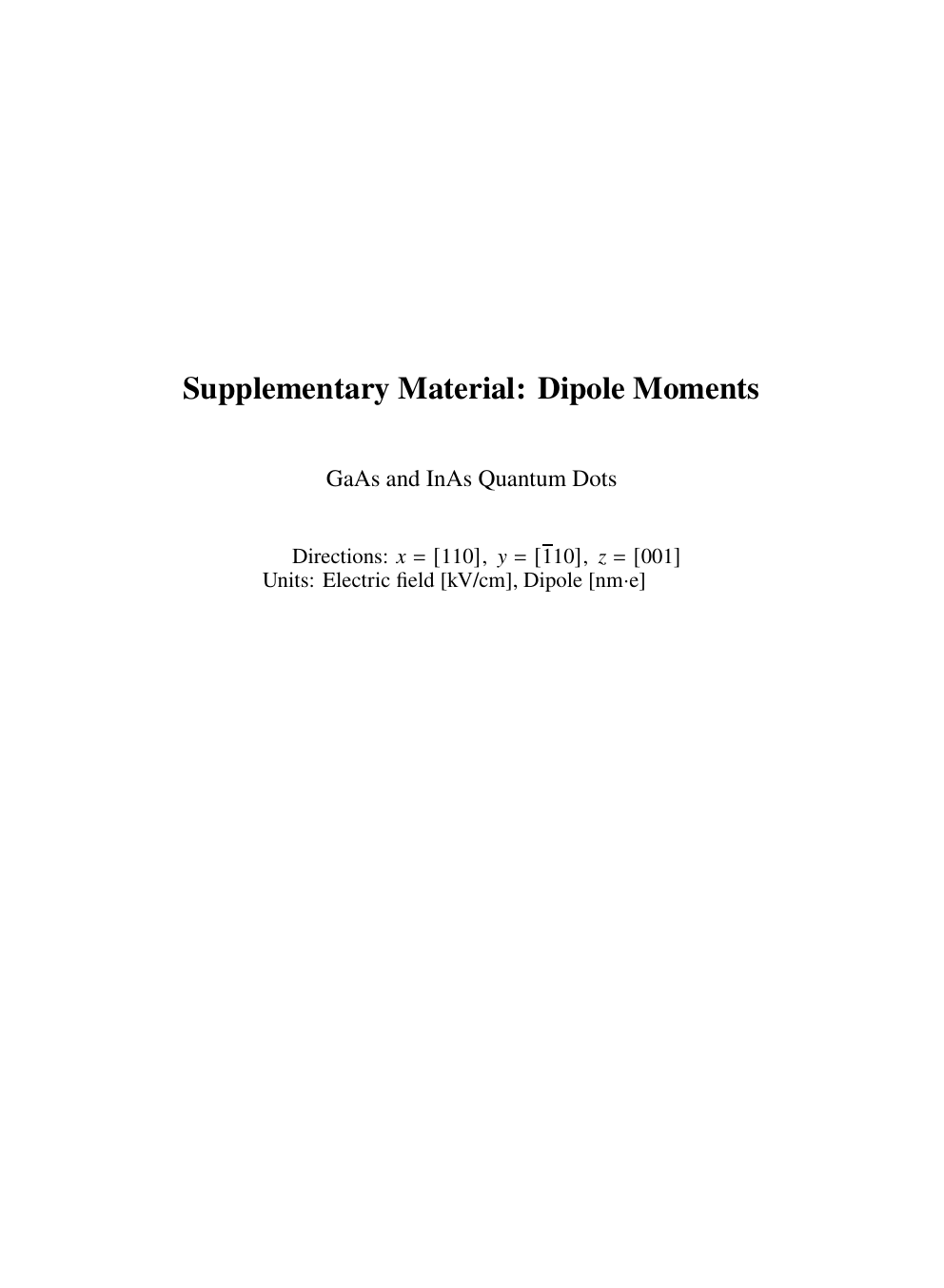}%
}

% ---------- wstawianie quadrupoli (3 strony) ----------
\foreach \i in {1,2,3} {%
  \clearpage
  \includepdf[pages={\i}]{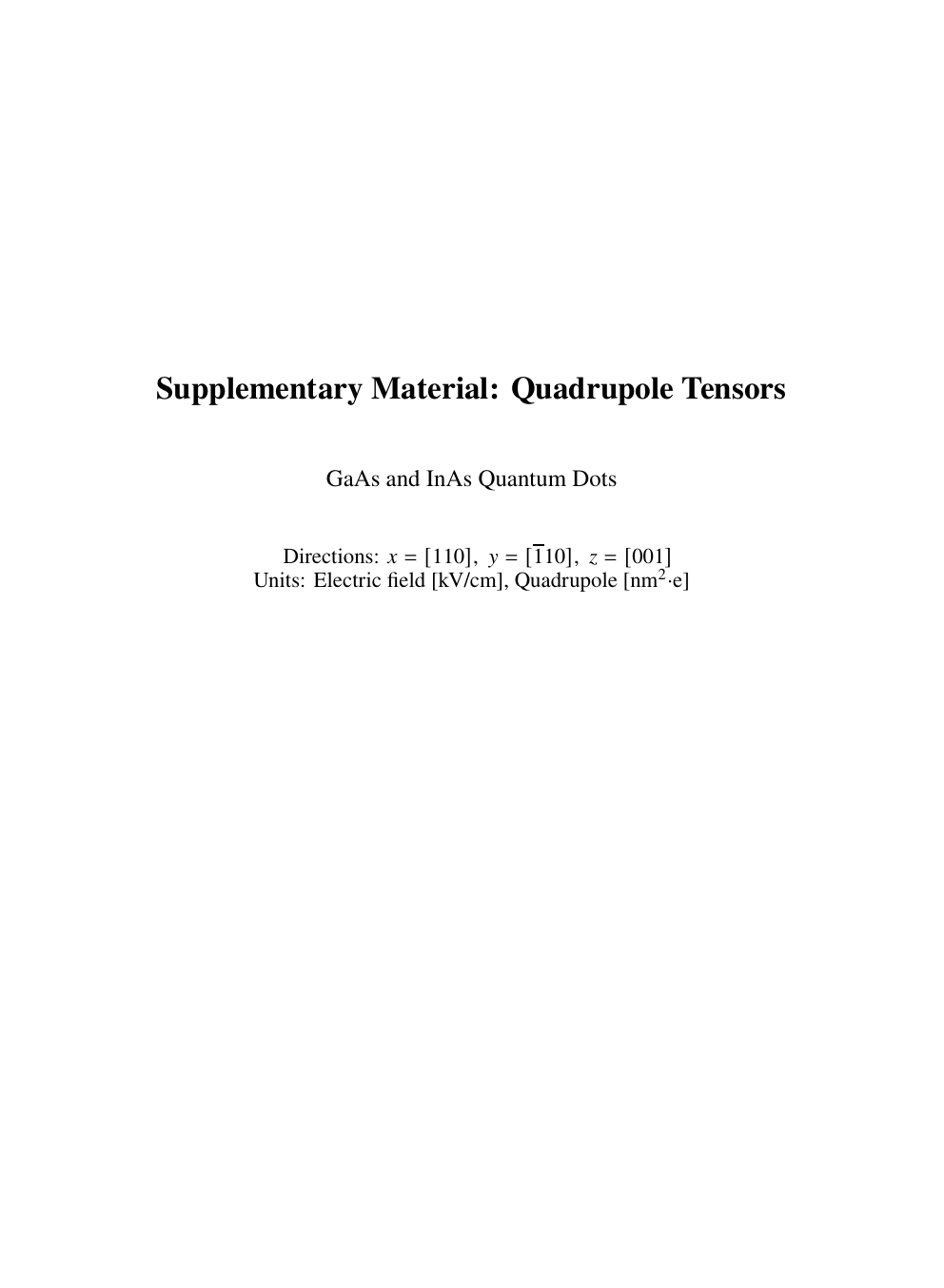}%
}
% \includepdf[pages=-]{Supplement/dipoles_tables.pdf}
% \includepdf[pages=-]{Supplement/quadrupoles_tables.pdf}

\end{document}